\documentclass[sn-nature]{sn-jnl}

\usepackage[utf8]{inputenc}
\usepackage{multibib}
\usepackage{soul}

\usepackage{graphicx}%
\usepackage{multirow}%
\usepackage{amsmath,amssymb,amsfonts}%
\usepackage{xltabular}
\usepackage{amsthm}%
\usepackage{mathrsfs}%
\usepackage[title]{appendix}%
\usepackage{xcolor}%
\usepackage{textcomp}%
\usepackage{manyfoot}%
\usepackage{booktabs}%
\usepackage{algorithm}%
\usepackage{algorithmicx}%
\usepackage{algpseudocode}%
\usepackage{listings}%
\usepackage{newtxtext,newtxmath}
\usepackage{colortbl}
\usepackage{lscape}
\usepackage{placeins}
\usepackage{wasysym}
\usepackage{bm}          
\usepackage{color}
\usepackage{xspace}
\usepackage{mathtools}
\usepackage{esdiff}
\usepackage{ulem}
\usepackage{wrapfig}
\usepackage{rotating}
\usepackage{xfrac}
\usepackage{relsize}
\usepackage{mathtools}
\usepackage{subcaption}
\usepackage{multirow}
\AtBeginEnvironment{longtable}{\tiny}
\usepackage{afterpage}
\usepackage{longtable}
\usepackage[T1]{fontenc}
\usepackage{ae,aecompl}

\newcommand{\Msun}{\ensuremath{\,\textrm{M}_\odot}\xspace}

\newcommand{\km}{{\ensuremath{\,\textrm{km}}}\xspace}

\definecolor{darkraspberry}{rgb}{0.53, 0.15, 0.34}
\definecolor{dodgerblue}{rgb}{0.12, 0.56, 1.0}
\definecolor{dogwoodrose}{rgb}{0.84, 0.09, 0.41}
\definecolor{cadmiumgreen}{rgb}{0.0, 0.42, 0.24}

\begin{document}

\title[Article Title]{Evidence for stellar mergers of evolved massive binaries: blue supergiants in the Large Magellanic Cloud}


\author*[1,2]{\fnm{Athira} \sur{Menon}}\email{athira.menon@iac.es}

\author[3]{\fnm{Andrea} \sur{Ercolino}}

\author[4]{\fnm{Miguel A.} \sur{Urbaneja}}

\author[1]{\fnm{Daniel J.} \sur{Lennon}}

\author[1,2]{\fnm{Artemio} \sur{Herrero}}

\author[5,6]{\fnm{Ryosuke} \sur{Hirai}}

\author[3,7]{\fnm{Norbert} \sur{Langer}}

\author[3]{\fnm{Abel} \sur{Schootemeijer}}

\author[8,9]{\fnm{Emmanouil} \sur{Chatzopoulos}}

\author[8]{\fnm{Juhan} \sur{Frank}}

\author[8]{\fnm{Sagiv} \sur{Shiber}}

\affil*[1]{Instituto de Astrofísica de Canarias, Avenida Vía Láctea s/n, 38205 La Laguna, Tenerife, Spain}

\affil*[2]{Universidad de La Laguna, Departamento de Astrofísica, Avenida Astrofísico Francisco Sánchez s/n, 38206 La Laguna, Tenerife, Spain}

\affil[3]{Argelander Institut für Astronomie,Auf dem Hügel 71, DE-53121, Bonn, Germany}

\affil[4]{Universität Innsbruck, Institut für Astro- und Teilchenphysik, Technikerstr. 25/8, 6020 Innsbruck, Austria}

\affil[5]{OzGrav: Australian Research Council Centre of Excellence for Gravitational Wave Discovery, Clayton, VIC 3800, Australia}

\affil[6]{School of Physics and Astronomy, Monash University, Clayton, Victoria 3800, Australia}

\affil[7]{Max-Planck-Institut für Radioastronomie, Auf dem Hügel 69, DE-53121, Bonn, Germany}

\affil[8]{Department of Physics \& Astronomy, Louisiana State University, Baton Rouge, LA, 70803, USA}

\affil[9]{Hearne Institute of Theoretical Physics, Louisiana State University, Baton Rouge, LA, 70803, USA}

\abstract{Blue supergiants are the brightest stars in their host galaxies and yet their evolutionary
status has been a long-standing problem in stellar astrophysics. In this pioneering work, we present a large sample of 59 early B-type supergiants in the Large Magellanic Cloud with newly derived stellar parameters and identify the signatures of stars born from binary mergers among them. We simulate novel 1D merger models of binaries consisting of supergiants with hydrogen-free cores (primaries) and main-sequence companions (secondaries) and consider the effects of interaction of the secondary with the core of the primary. We follow the evolution of the new-born 16--40\Msun stars until core-carbon depletion, close to their final pre-explosion structure. Unlike stars which are born alone, stars born from such stellar mergers are blue throughout their core helium-burning phase and reproduce the surface gravities and Hertzsprung-Russel diagram positions of most of our sample. This indicates that the observed blue supergiants are structurally similar to merger-born stars. Moreover, the large nitrogen-to-carbon and oxygen ratios, and helium enhancements exhibited by at least half our data sample are uniquely consistent with our model predictions, leading us to conclude that a large fraction of blue supergiants are indeed products of binary mergers.}

\maketitle

The evolution of massive stars (M$_\textrm{initial}>8$\Msun) has a strong impact on their host galaxies and are progenitors of supernovae, long gamma ray bursts and gravitational-wave systems. They are thus among the science drivers of large-scale surveys such as the upcoming Rubin-LSST which will observe millions of optical transients and ongoing LIGO-VIRGO-KAGRA gravitational-wave detectors, along with components of the future spectroscopic surveys such as WEAVE, 4MOST and SDSS-V/APOGEE. 

Due to their immense luminosities, blue supergiants (BSGs) are excellent spectroscopic calibrators of extragalactic distances and  interstellar extinction (e.g., \cite{kudritzki2012}), the galactic stellar mass-metallicity relation (e.g., \cite{bresolin2022}) and to trace galactic chemical evolution (e.g., \cite{kudritzki2016}).
However, their evolution is poorly understood.
For decades, the question of whether these BSGs are core-hydrogen burning stars in an extended
main sequence (MS), or core-helium burning stars that are either evolving redwards in the Hertzsprung-Russell 
diagram (HRD) or have looped back into the blue (performing a `blue loop') after a red supergiant (RSG) phase, has been debated. The answer will have a profound impact on our understanding of massive star and galaxy evolution. While clues as to their evolutionary status is provided by their stellar parameters and surface composition, 
single-star models have so far had limited success in explaining the overall observed population \cite{vink2010,georgy2021, bellinger2023, ma2023}.

About $50-70$\% of massive stars in our cosmic neighbourhood (the Milky Way and Magellanic Clouds) are observed to live in binary systems with close stellar companions \cite{sana2012,sana2013}. Binary evolution offers additional channels for the creation of BSGs, for example as mass gainers in binary systems \cite{farrell2019}. The low incidence of BSGs as components of binaries \cite{mcevoy2015,deburgos2023} however,  argues against this mechanism being a significant contributor to the population. 

By contrast, stellar merger models have had great success in explaining both the light curve of supernova SN~1987A which occurred in the Large Magellanic Cloud (LMC) and the properties of its BSG progenitor \cite{menon2017, menon2019,urushibata2018}, the B3\,I supergiant Sk-$69\,^{\circ}202$\xspace \cite{walborn1987}.
Based on the `slow merger' paradigm \cite{pods1990,ivanova2002c}, this model involves the merger of a core-He depleted red supergiant (RSG) with a main sequence (MS) companion via common envelope (CE), to produce a $20-24$\Msun core-C burning BSGs resembling Sk-$69\,^{\circ}202$\xspace that lives for $40-80$\,kyr before exploding.  The explosions of such post-merger models have also been able to reproduce the light curves of other Type-II peculiar SNe like SN~1987A \cite{menon2019}.

 Such `evolved' binary mergers, where the primaries are post-MS giants, have been evoked to explain luminous blue variables \cite{justham2014},  B[e] supergiants  \cite{pods2006}, supermassive stars  like $\eta$ Carinae \cite{hirai2021} and other supergiants observed with circumstellar rings and bipolar nebulae similar to Sk-$69\,^{\circ}202$\xspace, like Sher 25\cite{hendry2008,smith2007}. 



Inspired by the success of binary mergers in explaining SN~1987A and its progenitor, we build new binary merger models to confront a large sample of 59 B0\,I -- B3\,I stars in the LMC, for which we derive new stellar parameters and elemental abundances. We contrast our findings with single-star models and argue that stellar mergers are likely to be the largest contributors to the observed BSG population, and discuss some predictions of this scenario.

\section{Observational data}
\label{data}
The 59 B-type supergiants discussed here (the `BSG sample') are drawn from three sources; a `FEROS' dataset  (discussed by \cite{lennon2010A}), the Flames Survey of Massive Stars data (FSMS, \cite{evans2005, trundle2007, hunter2009}) and the VLT-FLAMES Tarantula Survey (VFTS, \cite{mcevoy2015,evans2011}).
For these sources we have derived stellar parameters, such as their effective temperatures (log\,T$_\textrm{eff}$), surface gravity (log\,g), surface luminosity (log\,L/L$_\odot$) and projected rotational velocity (vsini), along with the surface helium (He), carbon (C), nitrogen (N) and oxygen (O) abundances using the stellar atmosphere-wind code FASTWIND v10.6.4 \cite{puls2005}.
Further details may be found under sections~\ref{spectroscopic_data} and ~\ref{spectroscopic_analysis} , while results for individual stars are listed in Table~\ref{tab:abundances} under Extended Data.
FSMS also gives CNO abundances for MS O and B stars in the LMC \cite{hunter2009}, providing 
the LMC abundance baseline reference point for the present work. 

We express abundances as  $\epsilon\textrm{(X)} = \textrm{log}_{10}(N_\textrm{X}/N_\textrm{H}) + 12$, where $N_\textrm{X}$ and $N_\textrm{H}$ are the number fractions of the element and hydrogen respectively, except for helium, which is given as $Y = N_\textrm{He}/N_\textrm{H}$. 
We obtain log\,$\textrm{(N/C)}$ values from close to the LMC baseline to as high as 1.8\,dex, an enhancement factor of $\approx60$. BSGs classified as type BC by \cite{fitzpatrick1991} have close to baseline nitrogen abundances, consistent with the hypothesis that they are N-deficient relative to `morphologically normal' supergiants.
However, they are also oxygen rich relative to the LMC baseline by $\approx0.2$ to $-0.3$\,dex. Two possible explanations are: (i) some systematic abundance offsets between B-dwarfs (used to define the baseline) and BSGs; (ii) the baseline metallicity in the LMC may contain some 
scatter \cite{markova2020}.




Noteworthy for the present paper is the enhancement of helium we find in a third of the sources.  This is also in good agreement with the  recent work on Galactic  BSGs \cite{wessmayer2023}.



\section{Evolutionary Models}
To build our merger models, we first select precise binary models that are flagged to enter CE when the primary star
is a post-MS star and the secondary is a MS star. At this point the binary evolution is terminated. The selected binaries are thereafter merged via a detailed 1D process, and the post-merger star is evolved until core C-depletion, close to its pre-supernova state.
\label{models}

\subsection{Computational pipeline for binary mergers}
\label{pipeline}
 For our binary models, we use the `LMC Bonn grid' from \cite{marchant_thesis} (see section~\ref{physics_summary} for the main physics assumptions). Their initial  mass fractions are the LMC mixture of \cite{brott2011a}: $X_\textrm{H} = 0.7391$, $X_\textrm{He} =0.2562$ and Z (metallicity) = 0.0047 ($\approx\textrm{Z}_\odot/2$). Among these, `Case\,B' binaries that have primaries with He-rich cores dominate the grid, while `Case\,C'  binaries which have primaries with He-depleted cores form only 0.4\% of the grid.  We assume all these  CE-flagged models can  merge (see section~\ref{binary_grid} for CE termination conditions).

At the onset of the merger, the secondary star spirals-in towards the centre of the primary  due to viscous drag forces and may get tidally disrupted by the H-free core of the primary, resulting in a stream of H-rich secondary material to penetrate the core. In the meantime, the rest of the secondary dissolves in the envelope  and the merger is complete \cite{ivanova2002a, ivanovathesis}. To simulate our merger models, we adapt the above scenario to 1D based on the method of \cite{menon2017}, to the stellar evolution code \texttt{MESA} v.21.1 \cite{paxton2011,paxton2013, paxton2015,paxton2018,paxton2019}.

We consider the effect of penetration for two limits: at its maximum depth inside the H-free core ($M_\textrm{r, max}$) and just up to the boundary of the H-free core ($M_\textrm{r, H-free core}$). In both cases, the secondary is mixed uniformly down to the required depth. A first-order estimate of $M_\textrm{r, max}$ can be made by assuming the secondary stream  hits the core ballistically. We define the core dredge-up factor as $f_\textrm{c} = 1- \left(\displaystyle\frac{M_\textrm{r,\,penetration}}{M_\textrm{r, H-free core}}\right)$, where $M_\textrm{r,\,penetration}= M_\textrm{r,\,max}$ or  $M_\textrm{r,\,H-free core}$  (see section~\ref{1d_merger} for the derivation). 





\subsection{An exemplary Case\,B merger model}
\label{exemplary}
\begin{figure}
\centering
\includegraphics[width=\linewidth]{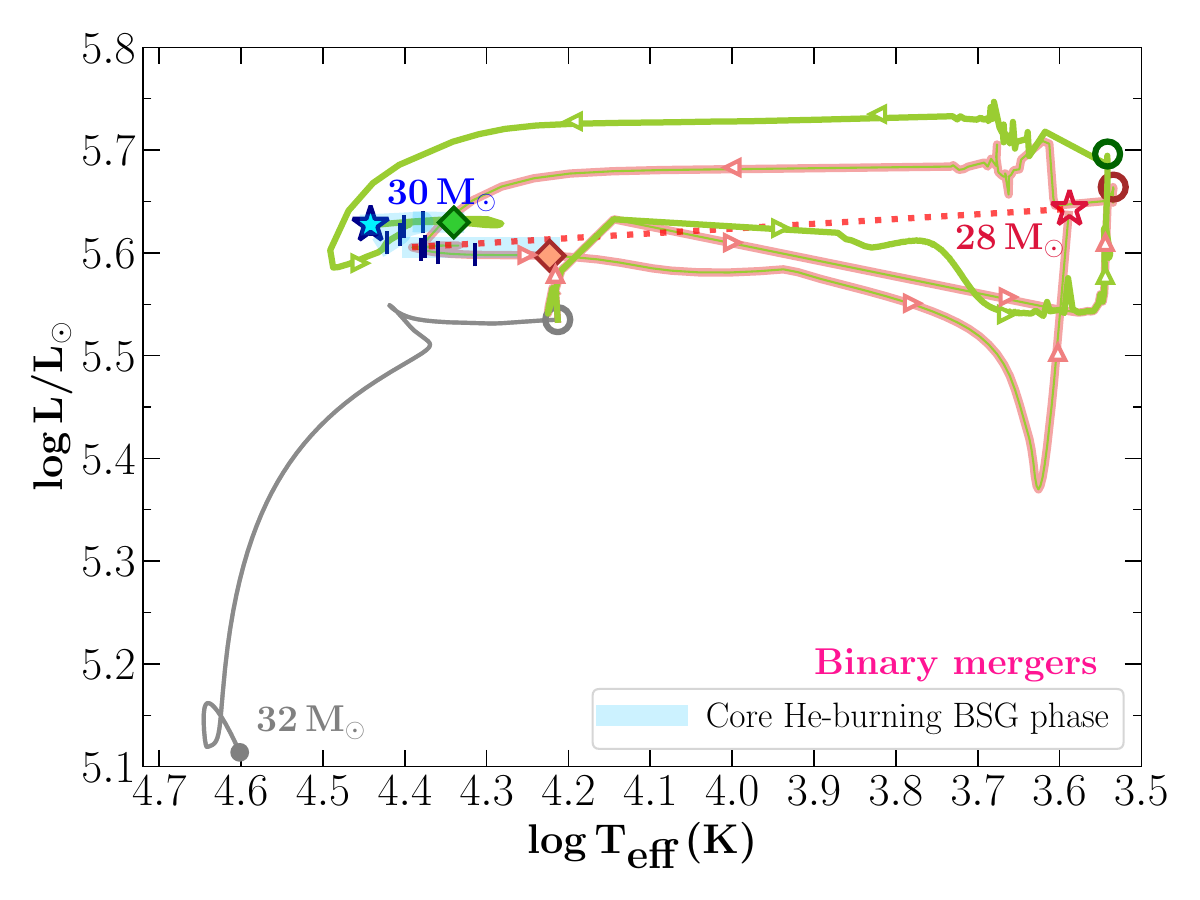}
  \caption{Evolutionary tracks of a 32\Msun (primary) + 3\Msun (secondary)  binary that merges when the primary is a 30\Msun star, with two core-penetration limits:  f$_\textrm{c,max}$ (system 1, green line) and  f$_\textrm{c} = 0$ (system 2, brown line). The direction of post-merger evolution is indicate by $\triangle$ symbols, along with tick (|) marks separated by 0.1\,Myrs during the core He-burning phase (light blue shading). Also plotted are the primary ZAMS position (grey $\bullet$), start (grey $\circ$) and end of merger (green and brown $\circ$), core He-depletion (filled $\Diamond$) and core-C depletion ($\star$), close to their respective pre-explosion BSG (blue $\star$) and RSG (red $\star$) position; the latter is connected by a red dotted line to its core He-burning phase. The secondary star is barely evolved from ZAMS and is not shown here.} 
\label{hrd_merger}
\end{figure}


In Fig.~\ref{hrd_merger}, we consider the 
merger of a binary model where both stars begin their evolution at the Zero Age Main Sequence (ZAMS) with an initial period of P$_\textrm{i} =  26.6$\,d. The initial mass of the primary is M$_\textrm{1,i}=32$\Msun and of the secondary is M$_\textrm{2,i}=3$\Msun. By nearly 6.2\,Myr, the primary star has lost $\approx2$\Msun due to winds and just begun core He-burning. Subsequently, it rapidly transfers mass at a rate of $\approx10^{-1}$\Msun/yr to its MS companion and undergoes unstable Roche lobe overflow, at which point a CE is expected to ensue.

The onward merger evolution is modelled for the two limits of stream-core interaction discussed in section~\ref{pipeline}: system 1 with $f_\textrm{c,max}=28$\%, and system 2 with $f_\textrm{c}=0$. The merger occurs over a duration of 300 years, after which the post-merger star has a total mass of 33\Msun. As the secondary mass is accreted with the same surface angular momentum as the primary, the immediate post-merger star in both systems spins up to critical velocity. Rotationally-enhanced mass loss quickly causes the star to shed $\approx3$\Msun and spin down to sub-critical equatorial velocity ($\approx100$\,km/s). As the homogeneously mixed envelope of the post-merger star has no molecular weight gradients, the star zooms towards the blue region of the HRD.


The post-merger stars of the two systems have different structures: that of system 1 has a reduced H-free core mass of M$_\textrm{H-free, core} \approx9$\Msun, a thick H-burning shell and a 21\Msun  H-rich envelope, while that of system 2 has an intact H-free core mass (M$_\textrm{H-free, core} \approx 12.5$\Msun), a thinner H-burning shell and a smaller envelope of 17.5\Msun.

Once core He-burning begins in gusto (log\,(L$_\textrm{He}$/L$_\odot)\gtrapprox4$), the post-merger star expands and loops back to cooler temperatures. Nevertheless in both cases, they retain their BSG nature throughout their core He-burning phase, which lasts for 0.45\,Myr for system 1 (due to its smaller H-free core mass) and 0.38\,Myr for system 2. As the  H-burning shell of the post-merger star of system 1 is more luminous than its He-burning core, it retains a hotter T$_\textrm{eff}$, while the reverse is true for system 2, causing the post-merger star to expand to cooler temperatures. 

The evolution after core He-burning depends on the luminosities of the H-burning shell and the He-burning shell, formed during core contraction after core He-burning. In system 2, the thin H-burning shell expands 
and gets eventually extinguished. With just a single shell-burning source (the He-burning shell), the post-merger star continues to expand to a 28\Msun RSG by the end of core C-burning, and subsequently explode as a Type II-plateau SN. On the other hand in system 1, the dominance of the H-burning shell persists. With two active shell-burning sources, the star undergoes an overall contraction and returns to the blue region of the HRD for core-C burning as a 30\Msun BSG,  subsequently exploding as  Type II-peculiar SN.

\begin{figure*}
\centering
\begin{subfigure}{0.495\linewidth}
  \centering\captionsetup{width=.95\linewidth}
  \includegraphics[width=\linewidth]{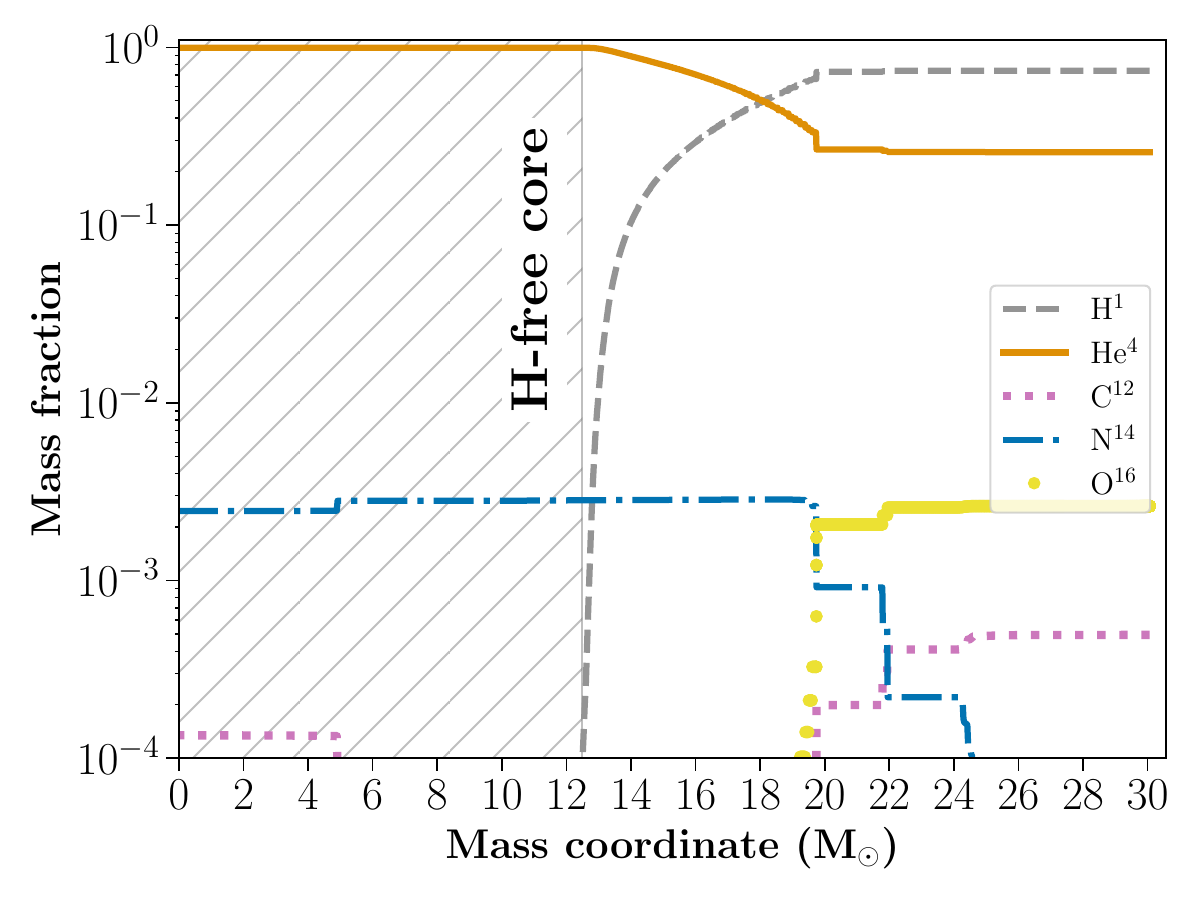}
\end{subfigure}
\begin{subfigure}{0.495\linewidth}
\centering\captionsetup{width=.95\linewidth}
  \includegraphics[width=\linewidth]{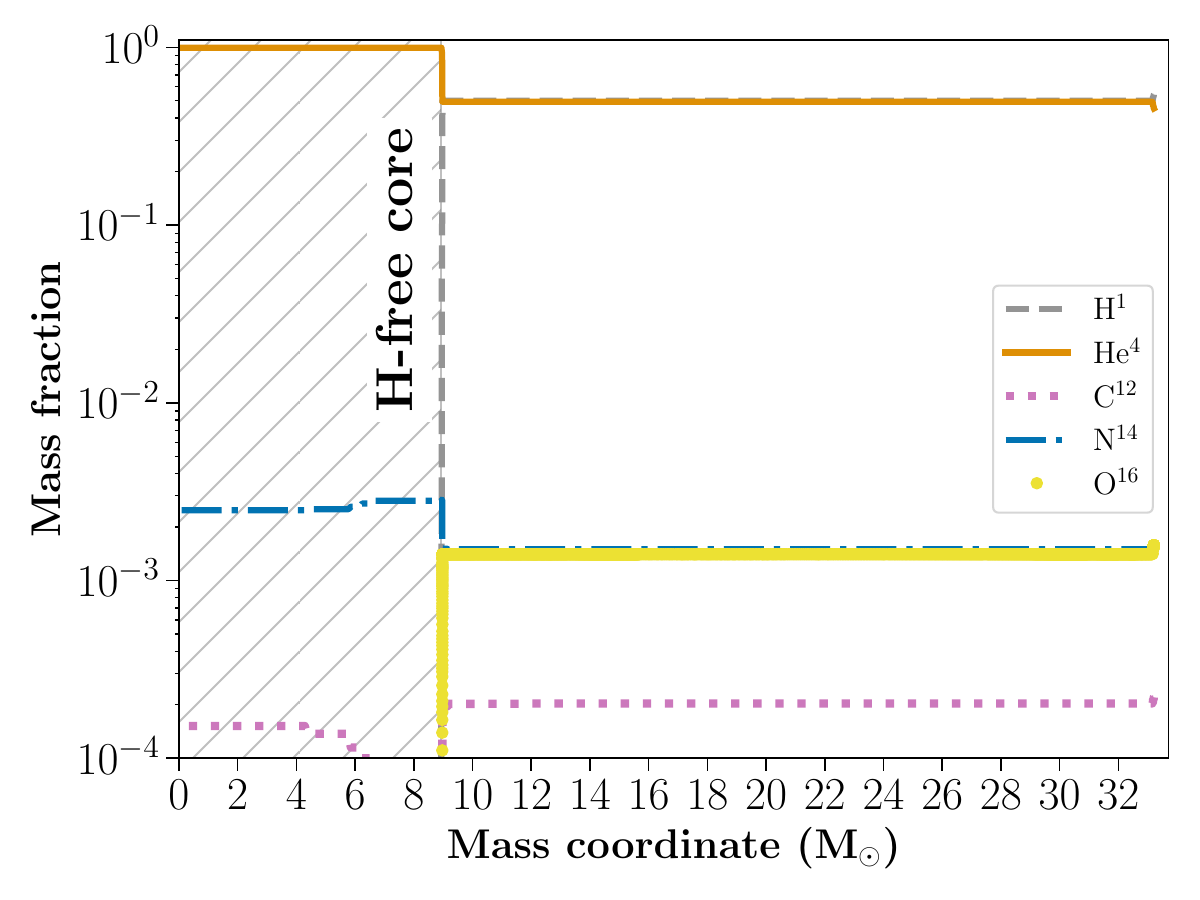}
\end{subfigure}%
\caption{Composition of the primary star at the onset of merger (left) and of the post-merger star at the end of the merger (right), corresponding to system 1 in section~\ref{exemplary}. \label{composition}}
\end{figure*}

The envelopes of these post-merger stars were mixed down to the H-burning shell of their primaries, leading to surface abundances close to CNO equilibrium. At the same time, carbon and oxygen are somewhat replenished due to the mixing of the MS secondary (see Fig.~\ref{composition} in section~\ref{1d_merger} for the composition profiles). As we shall see, these processes lead to a range of N/C and N/O ratios (and He abundances) which is not easily accessible by born-alone stars.



\subsection{The binary-merger models in this work}

Due to the stellar-wind and mass-transfer prescriptions in the Bonn grid (see section~\ref{binary_grid}), the Case\,C models are restricted to $10-11$\Msun core C-burning primaries, whose merger products directly inflate to $11-15$\Msun RSGs. Hence, we are unable to simulate the more massive Case\,C mergers required for the 20-24$\Msun$ progenitor of SN~1987A \cite{menon2017}.




In this study, we present the results of 47 Case\,B binary merger models with M$_\textrm{1,i}=10-40$\Msun, M$_\textrm{2,i}=1-32$\Msun and  P$_\textrm{i}=16-3162$\,days, which undergo core He-burning as $16-40$\Msun BSGs and are evolved until core C-depletion (the full parameter list of the models are in Table~\ref{tab:caseB}).



\subsection{Born-alone stars}
\label{born-alone}
We consider born-alone stellar models of initial masses of $16-40$\Msun, whose evolutionary tracks overlap with our observed sample on the HRD.   

Some of our sample BSGs coincide  with the HRD positions of  O-type MS stars. In the LMC, O-type MS stars are observed to have projected rotational velocities that peak at vsini$\approx80$\,km/s and tail off at $\approx200$\,km/s \cite{ramirez-agudelo2013}. We consider the evolution of single-star models with initial equatorial velocities of v$_\textrm{i,rot}\approx220$\,km/s from  \cite{brott2011a}, corresponding to the upper-limit contribution of rotating MS stars to our observed sample.




During core He-burning, single stars may perform blue loops as they return from the RSG to the BSG phase. These can be further categorized in to  ``classical" blue-loops, which typically occur in the low-luminosity regime (log\,L/L${_\odot}<5.4$) and the high-luminosity blue-loops. 

A combination of high semiconvection ($\alpha_\textrm{sc}$) and moderate overshooting efficiency parameters ($\alpha_\textrm{ov}=0.33$) were found to efficiently  produce  long-lived classical blue-loops in models of the Small Magellanic Cloud ($\approx\textrm{Z}_\odot/5$ metallicity)  \cite{schootemeijer2019}. We use the most effective combination of mixing parameters from \cite{schootemeijer2019} corresponding to  $\alpha_\textrm{sc}=100$ and  $\alpha_\textrm{ov}=0.33$ and build new LMC models with the same initial composition as the binary models.  Of these, only the 16\Msun and 20\Msun models perform classical blue-loops. 


High-luminosity blue loops occur due to wind-driven mass loss in their preceding RSG phase, causing stars to lose most of their H-rich envelopes before becoming BSGs.  We consider the \texttt{GENEVA} suite of LMC models \cite{eggenberger2021} (built with solar-scaled initial composition), among which the 32 and 40\Msun models perform high-luminosity blue loops that overlap with our sample T$_\textrm{eff}$ range. 

\section{Comparison between data and models}
To evaluate the evolutionary status of our sample BSGs, we simultaneously compare their surface abundances, HRD positions and surface gravities with the models. 

\subsection{Surface abundances}
\begin{figure*}
\centering
\includegraphics[width=\linewidth]{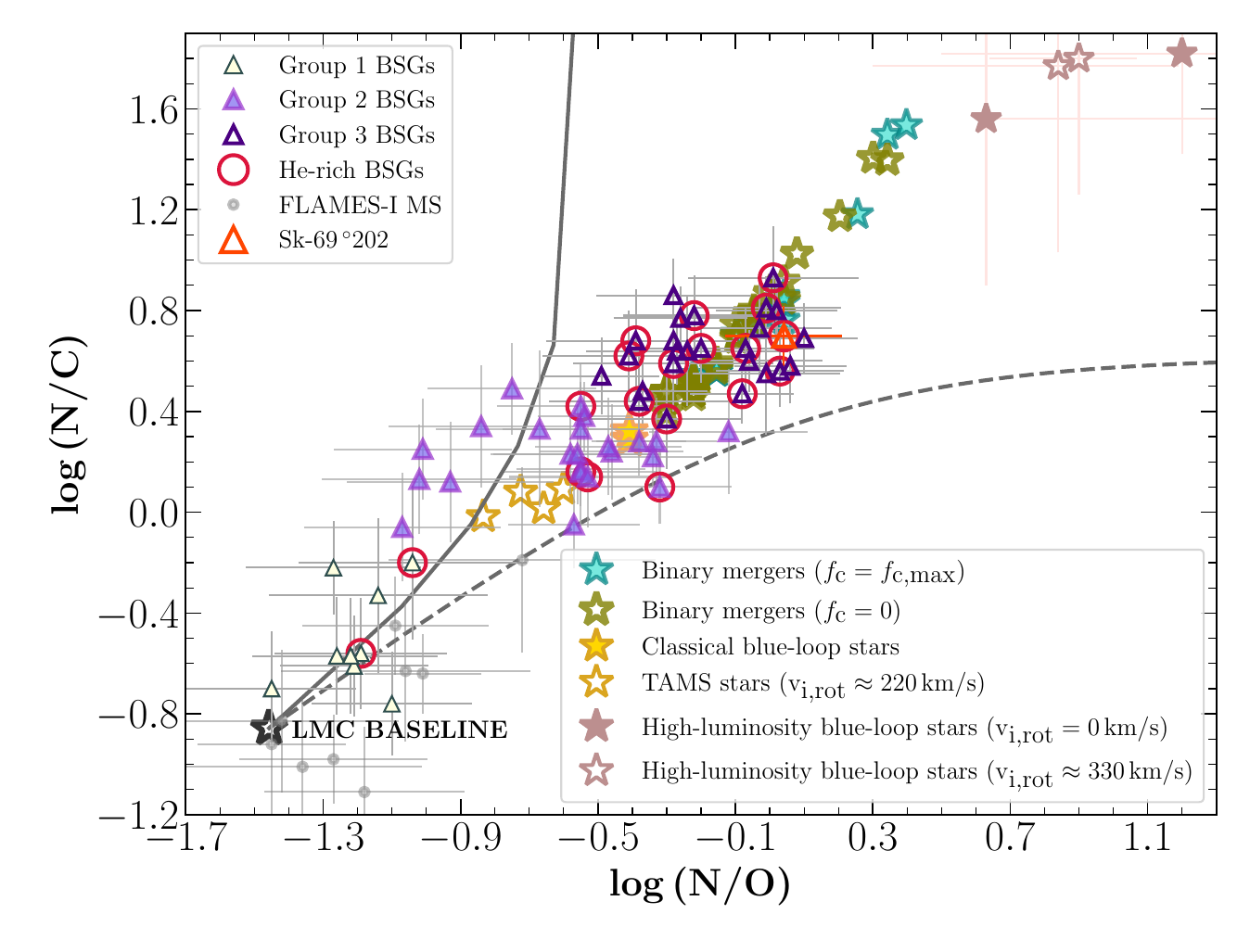}
    \caption{Log\, number ratios of N/C and N/O of our considered binary-merger and born-alone stellar models. The sample is marked according to the three groups as discussed in the text, along with the nebular abundances of the BSG progenitor of SN~1987A (dark orange $\triangle$, \cite{lundqvist1996}) and He-rich stars ($Y\geq0.12$). The grey dashed and solid lines depict the boundaries for the ON and CN cycle respectively using the analytical approximations of \cite{maeder2014}. \label{log_n_c_n_o}}
\end{figure*}

\begin{figure*}
\centering
\includegraphics[width=\textwidth]{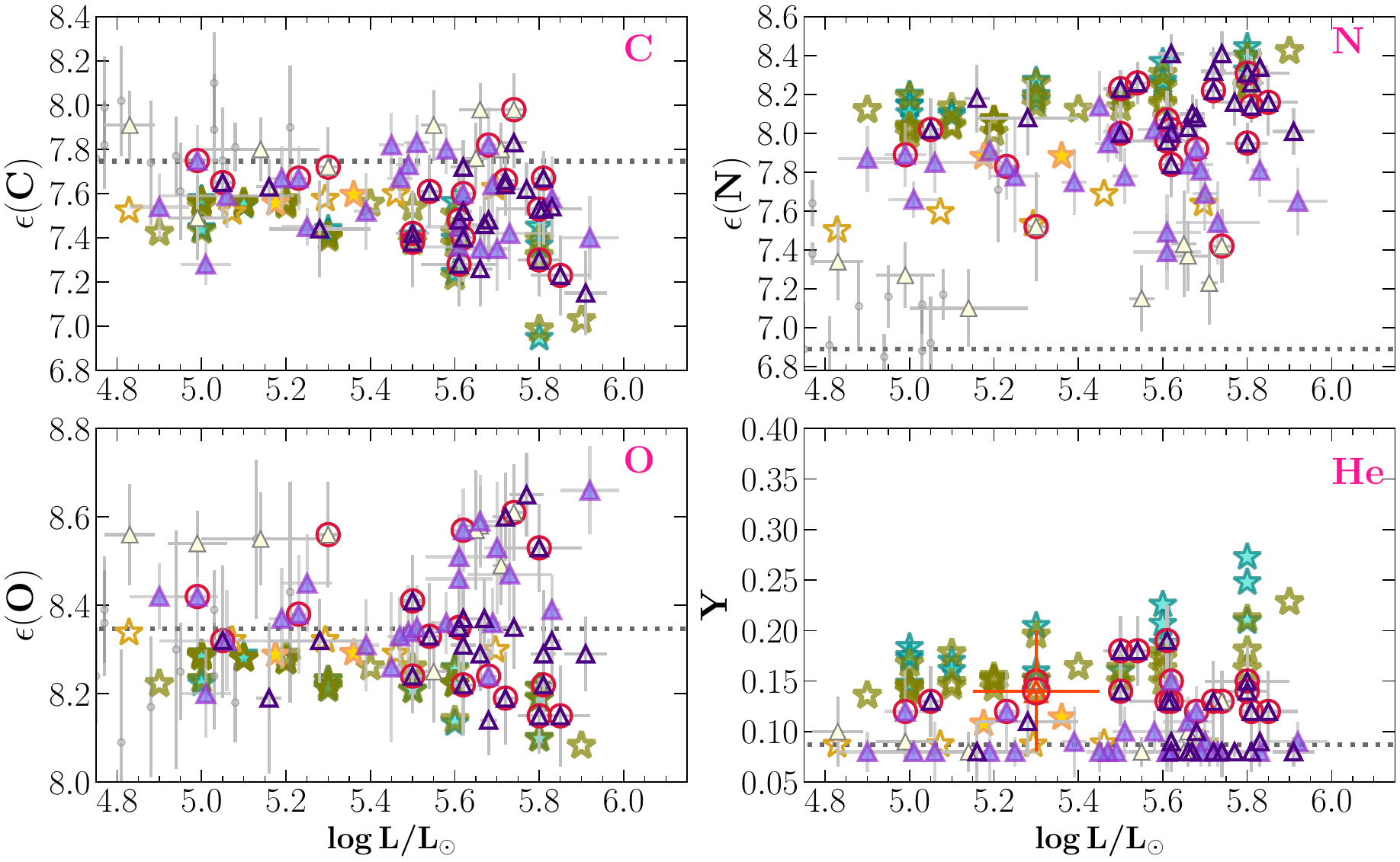}
\caption{Number fractions of C, N, O and He expressed  in log units as explained in section~\ref{data}, for merger models, classical blue- loops and TAMS stars, along with our sample groups (as plotted in Fig.~\ref{log_n_c_n_o}). $Y$ abundance of Sk-$69\,^{\circ}202$ is taken from \cite{france2011}. Dotted lines in each panel represent the initial abundances:  $\epsilon\textrm{(C,i)}= 7.75$, $\epsilon\textrm{(N,i)}= 6.90$, $\epsilon\textrm{(O,i)}= 8.35$ and $Y_\textrm{i}=0.087$.}
\label{surf_bsgs_all}
\end{figure*}
Guided by the N/C and N/O number ratios of the models, we broadly split our observed sample into three groups (within the error-bars of the sample), as indicated in Fig.~\ref{log_n_c_n_o}:

\begin{enumerate}
    \item \textbf{Group 1} ($\textrm{log\,(N/C)}< -0.2$, $\textrm{log\,(N/O)}\lessapprox-1$): containing 9 stars whose number ratios are explained by single-star models with v$_\textrm{i,rot}\lessapprox220$\km/s. 
    
    \item \textbf{Group 2} ($-0.2\leq\textrm{log\,(N/C)}\lessapprox0.35 $, \textrm{$-1<\textrm{log\,(N/O)}\lessapprox-0.25$)}: is a mixed group of 23 stars whose number ratios can be explained with classical blue-loop models or terminal-age MS (TAMS) models with  v$_\textrm{i,rot}\geq220$\km/s or even  merger models, as we shall in this section.

    \item \textbf{Group 3} ($\textrm{log\,(N/C)}\gtrapprox0.35$, $\textrm{log\,(N/O)}\gtrapprox-0.25$): contains 25 stars whose number ratios are  explained by merger models, but not by the born-alone stellar models.
\end{enumerate}

From Fig.~\ref{log_n_c_n_o}, we see that born-alone stars may explain the N/C and N/O ratios 
of the group 1 sample as MS stars, and also of the group 2 sample as classical-blue loop stars or as TAMS stars with fast v$_\textrm{i,rot}$. The high-luminosity blue-loop models have extreme N/C and N/O ratios that are far larger than the observed range, as these models have been stripped off most of their H-rich envelopes. Thus these high-luminosity blue-loop models do not explain the abundances of our sample.  


The group 3 sample which contains stars with the highest N/C and N/O ratios are not explained by blue-loop or MS stars, and are uniquely replicated by merger-born stars. This is evident when examining individual elemental abundances in Fig.~\ref{surf_bsgs_all}: mergers result in significant nitrogen enhancements (1.1 to 1.7\,dex) and carbon and oxygen depletions (up to $\approx0.75$ and 0.25\,dex, respectively), while classical blue loops or rotational mixing only lead to moderate nitrogen enhancements ($\lessapprox1$\,dex) and minor carbon and oxygen depletions relative to mergers.

It is worth noting that potential systematic offsets in the data (discussed in section~\ref{data}), may cause some stars with moderate log(N/C) ratios to lie outside the ON equilibrium line in Fig.~\ref{log_n_c_n_o}, due to their measured oxygen abundance being higher than the baseline. This offset could lead to an increase in the N/O ratio for some stars in group 2, aligning them with group 3, which is exclusively explained by mergers.




About one-third of the sample displays helium enrichment by a factor of 1.6 to 2.5 times higher than baseline, which is not only consistent with the helium abundances of merger models but also with Sk-$69\,^{\circ}202$\xspace \cite{france2011}. These He-rich BSGs are predominantly found in group 3, with a few also in group 2. Such helium enhancements are not produced by fast-rotating  or classical blue-loop born-alone stellar models. High-luminosity blue-loop models tend to exceed the observed range of helium enhancements.



Some stars in the sample are substantially enriched in nitrogen ($\geq1$\,dex) but only have baseline helium. This pattern is challenging to explain with any evolutionary scenario.However these stars are predominantly cool (T$_\textrm{eff}\lesssim$\,20,000\,K) and their He abundance depends primarily on the He\,I lines, offering rather weak diagnostics for the He abundance. The suspicion therefore arises that their He abundance and hence, the number of He-rich BSGs in this sample, are underestimated. Considering these uncertainties, we may infer that over half of our sample aligns with the abundance predictions of our merge models.

 \subsection{HRD positions and surface gravities}

\begin{figure*}
\centering
\begin{subfigure}{0.95\linewidth}
  \centering\captionsetup{width=\linewidth}
  \includegraphics[width=0.8\linewidth]{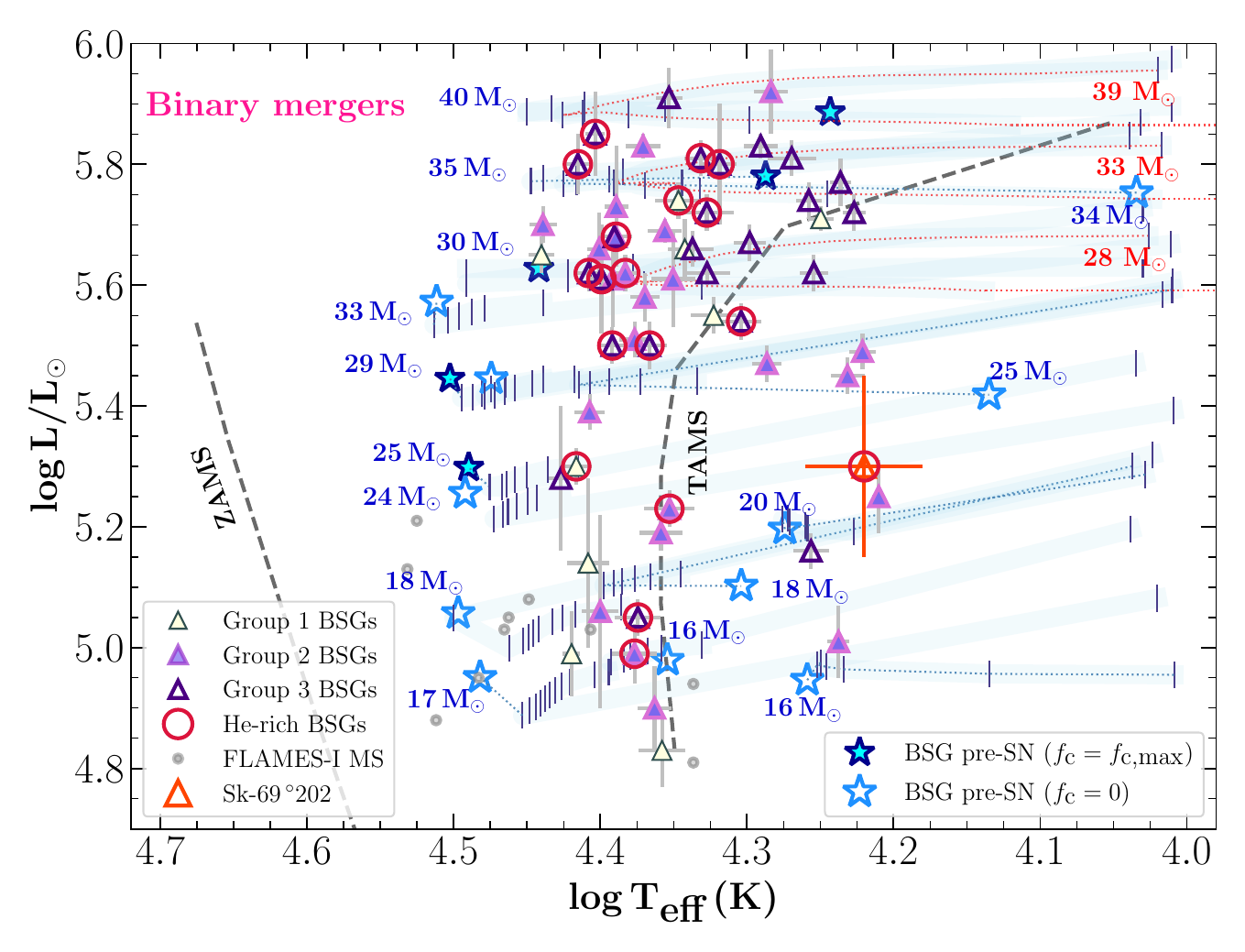}\vspace{-5pt}
  \caption{Core He-burning BSG phase of our merger models (light blue shading) and their corresponding SN progenitors ($\star$ symbols) and their  masses at core C-depletion, as in  Fig.~\ref{exemplary}. For BSG progenitors with log\,T$_\textrm{eff}\lessapprox4.3$ and RSG progenitors (red, log\,T$_\textrm{eff}<4.0$), we connect their full HRD trajectory. Errors of MS sample are about 0.1\,dex for both log L/L$_\odot$ and log\,T$_\textrm{eff}$.}
\end{subfigure}
\vspace{-10pt}
\begin{subfigure}{0.95\linewidth}
\centering\captionsetup{width=\linewidth}
  \includegraphics[width=0.85\linewidth]{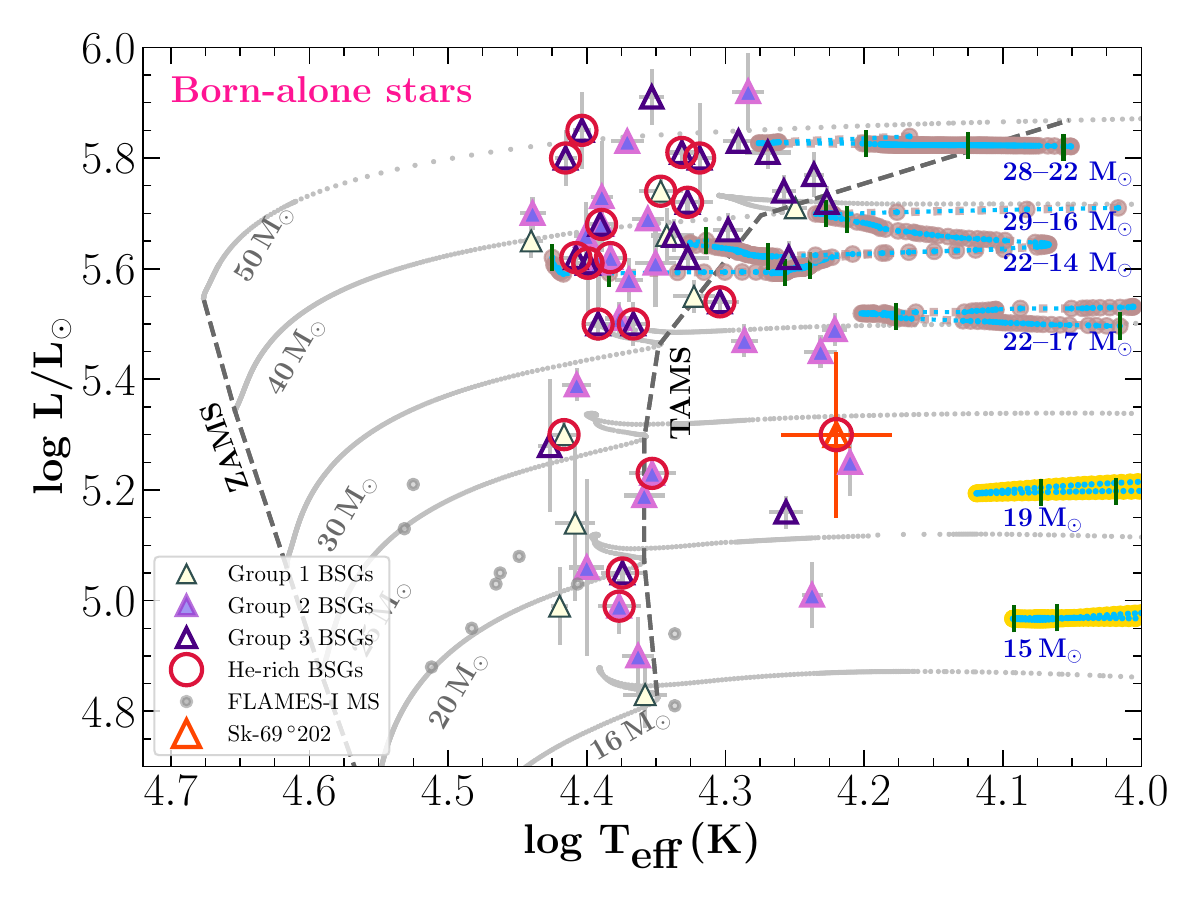}\vspace{-10pt}
  \caption{Blue-loop phase (blue dotted lines) of the classical (yellow tracks) and high-luminosity (light brown lines) categories, along with their mass in this phase. Background grey evolutionary tracks are of single-star models with v$_\textrm{i,rot}\approx220$\,km/s from \cite{brott2011a}.}
\end{subfigure}\vspace{5pt}
\caption{HRD showing the BSG phase of merger-born (top) and born-alone stars (bottom) and our sample (as in Fig.~\ref{log_n_c_n_o}), along with the position of Sk-$69\,^{\circ}202$  \cite{woosley1988}.  Tick marks indicate intervals of 0.1\,Myr in time in the BSG phase. For reference, the ZAMS and TAMS curves of v$_\textrm{i, rot}\approx220$\,km/s single star models \cite{brott2011a} are also plotted. \label{HRD}}
\end{figure*}

\begin{figure*}
\centering
\includegraphics[width=0.9\linewidth] {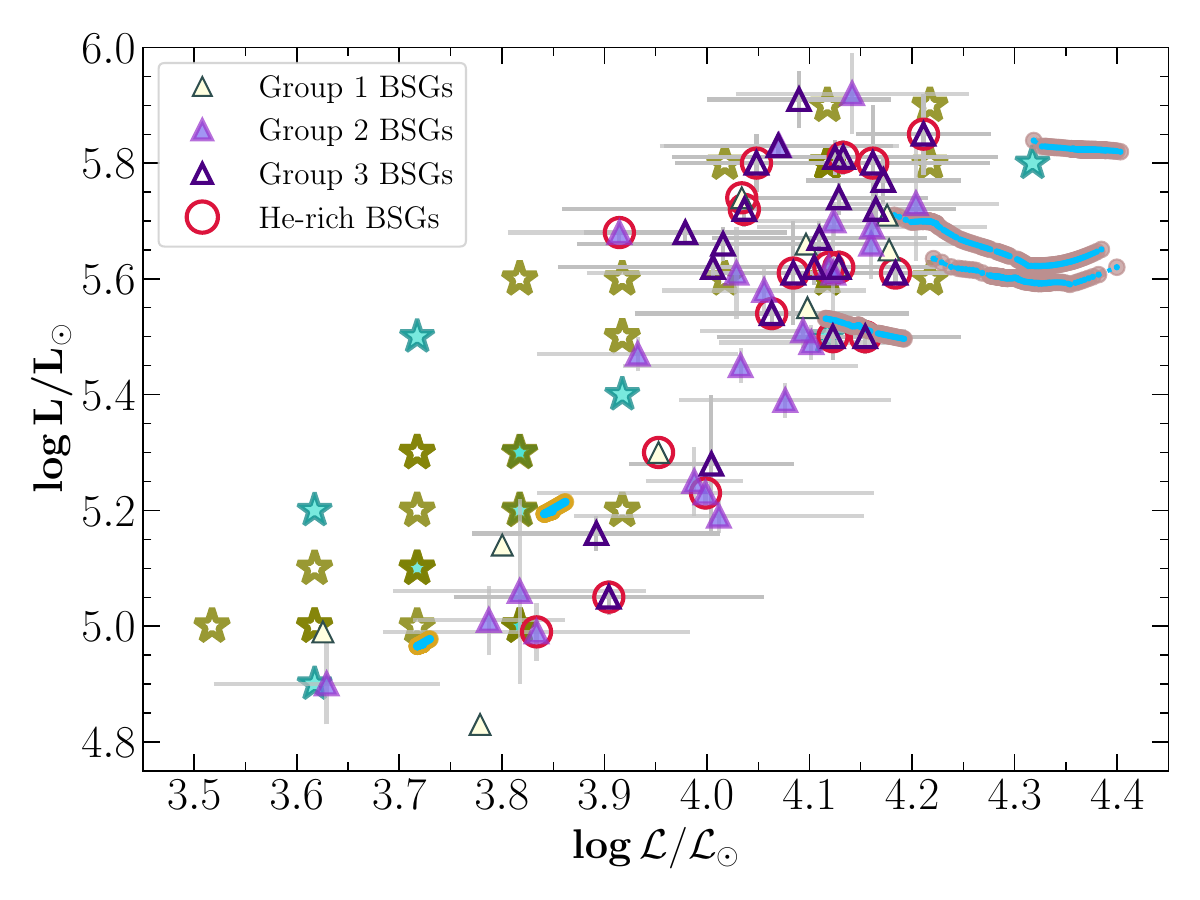}
\caption{Distribution of post-merger and blue-loop single-star models, along with the observed sample in the bolometric luminosity (log\,L/L$_\odot$) and spectroscopic luminosity space (log\,$\mathcal{L}$/$\mathcal{L}_\odot$). Symbols are as explained in Figs.~\ref{log_n_c_n_o} and \ref{HRD}. \label{L_Lspec}}
\end{figure*}

We next consider the disposition of the above groups in the HRD in Fig.~\ref{HRD}. Our $16-40$\Msun post-merger stars are BSGs throughout their core He-burning phase, spanning $1.6-0.6$\,Myrs,  and covering nearly the entire observed HRD. The blue-loop models, on the other hand, only last for $\lessapprox50$\% of their core He-burning lifetime and mostly fall short of log\,T$_\textrm{eff}\geq4.3$\,(K), where the majority of the sample lies. Thus the high-luminosity sample (log\,L/L$_{\odot}\geq5.4$), where most of the group 3 members also lie, is exclusively explained by the merger models.

In the lower-luminosity regime, covered by models under $\lessapprox30$\Msun, the classical blue-loops of single stars only extend up to log\,T$_\textrm{eff}\approx4.1$\,(K) and fail to overlap with the observed BSGs. In contrast, 
merger models reproduce this low-luminosity domain, which also includes the group 2 stars. Group 1 BSGs coincide with the FSMS-MS sample on the HRD and in their abundances; we thus infer that these BSGs may be MS stars.


Insight into the internal structure and the current mass of the observed BSGs can be obtained from their surface gravities. Rather than comparing the
gravities directly, which depend sensitively on the stellar radius,
we consider their spectroscopic luminosity $\mathcal{L}=T_\textrm{eff}^4/g$
in Fig.\,\ref{L_Lspec}, which is proportional to $L/M$. Because
massive post-MS single stars largely evolve at constant luminosity proportionally to their mass, the HRD locations of their BSG models in Fig.\,\ref{L_Lspec}, move further to the right the more mass they lose. This is particularly useful to distinguish whether the observed stars have massive H-rich envelopes or have lost their  H-rich envelopes.

In the lower-luminosity regime (log\,L/L$_{\odot}<5.4$) corresponding to blue-loop models with initial masses $\lessapprox 30\,$M$_{\odot}$, we see that mass loss is not significant as the positions of the observed stars agree well with our merger models. The indicated high-luminosity single-star models, however, only return to the BSG region after having lost most of their H-rich envelope and do not match the observations. Notably, because of the strong mass loss, these models also lost copious amounts of angular momentum, and fall short of recovering the observed rotation rates of the BSGs (Table~\ref{tab:abundances}) by one order of magnitude. Therefore in the high-luminosity regime also, the observed stars agree with the positions of the merger models. Overall thus, we may conclude that the observed BSGs are core He-burning stars with homogeneously mixed, massive H-rich envelopes similar to our merger models.




Finally, a few of the observed high-luminosity BSGs with log\,T$_\textrm{eff}\lessapprox4.3$\,(K) are located on the HRD in the region where Sk-$69\,^{\circ}202$ is also located. These stars may also be explained with massive Case\,C merger models \cite{menon2017}. Indeed, their N/C and N/O ratios are also consistent with the predictions from these models.


\label{comparison}

\section{Conclusions}
\label{conclusions}
Our work shows how Case\,B binary-mergers lead to BSGs whose surface parameters,  core-envelope structures and final fates are distinctly different from those born alone. We have found substantial evidence for imprints of such binary mergers in our  observational sample of nearly 60 BSGs in the LMC. The HRD positions and surface gravities of most of the sample are successfully reproduced by our merger models, indicating that early B-type supergiants structurally resemble 16--40\Msun stars with homogeneous massive H-rich envelopes that will sustain their BSG nature throughout their core He-burning lives. Combined with the possibility of  spin-up during the merger process and contrasting with available single-star models,  we conclude that a large fraction of early B-type supergiants, and potentially other classes of BSGs, are likely to be the results of  evolved (Case\,B or C) stellar mergers.


The helium enrichments of a third of the sample and  the high N/C and N/O ratios of over half the sample are uniquely explained by our merger models, and are also consistent with the measured nuclear abundances of the BSG progenitor of SN~1987A.   While the fate of our merger-born stars with $\textrm{M}<30$\Msun is to explode as Type II-peculiar SNe (like SN~1987A), the more massive ones may also explode as Type II-plateau SNe.

Our work thus identifies key imprints of massive-binary mergers that distinguishes them from born-alone stars, and paves the way towards exploring their contributions to the resolved stellar populations and core-collapse supernovae of galaxies in the Local Group and beyond.

\backmatter

\bmhead{Acknowledgments}
AM, DL and AH are supported by the Spanish Government Ministerio de Ciencia e Innovación through grants PID2021-122397NB-C21 and SEV 2015-0548. AM also acknowledges the Juan de la Cierva Incorporación (2022) fellowship. AE is supported by the DFG through grant LA 587/22-1. AM also thanks Alexander Heger for supporting the development of this work.

\section{Methods}
\label{methods}

\subsection{Spectroscopic Data}
\label{spectroscopic_data}
As noted, the spectroscopic data for the BSG sample is drawn from three sources, designated FEROS, FSMS and VFTS in the main text. 
The FEROS dataset is a high resolution ($R\sim48\,000$) survey of B-type supergiants in the LMC performed using the FEROS echelle spectrograph on the ESO 2.2m telescope, the data covering the wavelength region 3500--9200\,\AA.  
The FSMS \cite{evans2005} and VFTS \cite{evans2011} data are taken from two large programs that used the Flames multi-object spectrograph on the ESO VLT to observe OB stars in the Magellanic Clouds, the former being a high resolution survey ($R\sim 20\,000$) of OB stars in both Clouds, covering regions 3950--4750\,\AA\ and 6300--6700\AA, while latter was an intermediate resolution ($R\sim 7\,000$) survey of OB stars in the Tarantula Nebula of the LMC covering regions 3960--5070\,\AA\ and 6400--6820\,\AA.
For these latter two wide-ranging surveys we restrict the sample to apparently single stars with spectral types between B0 and B3 inclusive, the range of validity of the spectroscopic analysis, and luminosity class I.  A small number of variable and peculiar objects were excluded, as were a several hypergiants with very high mass-loss rates.  All the spectra considered here have signal-to-noise ratios of approximately 200.

One should note that this work is focused on early to mid B-type supergiants, hence the absence of late-B and A/F-type supergiants in the sample. In fact there are very few BSGs in this cooler temperature range with published CNO abundances, albeit \cite{hill1995} finds a mean under-abundance in [C/H] of only $\sim$0.2\,dex for a sample of 9 F-type supergiants in the LMC using an LTE analysis. 

\subsection{Spectroscopic Analysis}
\label{spectroscopic_analysis}

The FSMS and VFTS data have been reported on extensively in the literature, that included reports on the surface composition the B-type supergiant populations \cite{trundle2007,hunter2009,mcevoy2015}.
Additionally, preliminary results for the FEROS dataset are also available \cite{lennon2010A}.
However these results are based on an analysis using a TLUSTY \cite{hubeny1995} plane parallel NLTE model atmosphere grid \cite{dufton2005} that adopted a normal helium abundance, and did not take account of the impact of the stellar wind on the spectrum.
While the effect of neglecting the wind may have only moderate impact on the derived CNO abundances \cite{dufton2005}, the helium abundance provides important additional critical diagnostic value.
Furthermore, the VFTS results \cite{mcevoy2015} focus solely on the abundance of nitrogen.
Therefore we have re-determined the stellar parameters and surface HeCNO abundances using FASTWIND v10.6.4 \cite{santolayarey1997,puls2005} that accounts explicitly for the presence of a stellar wind.
The approach was broadly similar in methodology to that employed by \cite{urbaneja2005,urbaneja2017}, with the exception that the C\,{\sc ii} model atom of \cite{carneiro2019} was adopted, and has been widely used in the determination of the surface composition of O-type stars and B-type supergiants in Local Group Galaxies\cite{bresolin2022,grin2017,urbaneja2005}.
For our FSMS subsample of 13 objects we find good agreement with published abundances \cite{hunter2009}, deriving mean CNO abundances of 7.69$\pm$0.11, 7.82$\pm$0.32 and 8.39$\pm$0.12 compared with 7.62$\pm0.20$, 7.70$\pm$0.38 and 8.36$\pm$0.09.
Similarly, for our sample VFTS sources we derive a mean nitrogen abundance of 7.72$\pm$0.33, compared with a mean value of 7.70$\pm$0.38 for the same 14 sources \cite{mcevoy2015}.
Results are listed in Table 1, along with projected rotational velocities (vsini) previously published for the VFTS and FSMS subsamples. In addition we list vsini measurements for the FEROS subsample, obtained using the same Fourier Transform technique \cite{sergio2007}.

\subsection{Stellar physics assumptions} 
\label{physics_summary}
We summarise the main \texttt{MESA} input parameters used in all models presented in this paper. Further details about their assumptions and justifications are available in \cite{marchant_thesis}.

For convection, we use the Ledoux criterion and a mixing length parameter of $\alpha_\text{MLT}=1.5$ and a semiconvection efficiency parameter of  $\alpha_\text{sc}=0.01$. The efficiencies of other mixing processes are implemented as follows: thermohaline mixing with $\alpha_\text{th}=1$,  H-burning core overshooting  with $\alpha_\textrm{ov}=0.335$ of the convective boundary pressure scale height,  rotational mixing efficiency of $f_\text{c}=1/30$ and an efficiency parameter $f_\mu=0.1$ against molecular weight gradients.

The adopted wind mass-loss recipes are based on the $T_\textrm{eff}$ being hotter or cooler than the bi-stability jump temperature,  $T_\textrm{jump}\approx25\,\textrm{kK}$ \cite{vink2001}. In the cold regime, the maximum mass-loss rate between the \cite{vink2001} and \cite{nieu1990} is used while in the `hot' regime, the recipe depends on the surface hydrogen abundance ($X$): the \cite{vink2001} recipe for $X>0.7$, the \cite{hamann1995} recipe  (reduced by a factor of 10) for $X<0.4$ and for  $0.7>X>0.4$,  a linearly interpolated rate between the two `hot' recipes. 
 
The nuclear network used is initially set to the \texttt{basic} net during the MS phase and  extended to include $\alpha$-capture chains for later stages (the synthetic \texttt{approx21} network).

\subsection{The binary grid}
\label{binary_grid}
The LMC Bonn grid is a series of binary star models computed with \texttt{MESA} (Modules for experimental astrophysics) r8118 
\cite{paxton2011,paxton2013,paxton2015}, in which both stars are evolved simultaneously over an initial parameter space of primary masses M$_\textrm{1,i}=10,\ldots,40\,\textrm{M}_\odot$, mass ratio q$_\textrm{i}=\frac{\textrm{M}_\textrm{2,i}}{\textrm{M}_\textrm{1,i}}=0.10,\ldots,0.95$ and orbital period P$_\textrm{i}=1.4,\ldots,3162.3\,\textrm{d}$ \cite{marchant_thesis}.  These models are computed using the `Roche-lobe' binary mass-transfer scheme and assumed to be tidally synchronized at the zero age main sequence  (ZAMS).

Each binary model is initiated with both stars on the ZAMS and terminated either at core carbon(C)-depletion or when a relevant criterion for CE is met, such as: if the model reaches the maximum mass-loss rate that can be expelled via radiation from a disk surrounding the accretor, or, when the mass-transfer rate reaches $0.1\,\textrm{M}_\odot\textrm{yr}^{-1}$ (which is higher than thermal-timescale mass transfer) or, at the first instance of mass transfer in systems with  q$_\textrm{i}\leq \textrm{q}_\textrm{crit} = 0.25$, which is expected to be dynamically unstable. 

Case\,B and Case\,C mass transfer is highly inefficient  since a small amount ($\lesssim10$\%) of accreted mass is sufficient to spin up the secondary to critical rotation. Hence most of the transferred mass is not accreted by the secondary but is instead assumed to be lost from the system carrying the same specific orbital angular momentum as the accretor.

The parameter space for late Case\,B and Case\,C mass transfer is affected by several physics assumptions and their implementations. Our current findings are for a rather inefficient semiconvection parameter of $\alpha_\textrm{sc}=0.01$. The wind mass-loss rates may also be too high for RSGs, which in turn affect their radius evolution and binary separations. The `Roche-lobe' mass-transfer scheme used prevents the deep convective envelope of RSGs from expanding during mass transfer and causes convergence failures in some wide Case\,C systems of the grid.  Including an extended atmosphere or pulsations may also widen the Case\,C parameter space \cite{ercolino2023}.

\subsection{Our 1D merger method}
\label{1d_merger}

In our 1D merger approximation, we  take the exact structures of the post-MS primary and MS secondary stars when the binary model is flagged to enter a CE phase. We further assume that the MS star is always tidally disrupted by the H-free core of the primary as it spirals in through the CE, before eventually merging with it \cite{ivanovathesis}.

To simulate this in 1D, we accrete the averaged composition of the secondary mass on the primary at a rate of 0.01\Msun/yr and with the same entropy and angular momentum as its surface, while simultaneously mixing it up to a certain defined boundary inside the primary. By the end of the merger, the resultant star has a chemically homogeneous envelope and a He core mass determined by the penetration depth of the secondary. Although in reality this is a 3D process, it is expected that the penetration and the subsequent mixing occurs rapidly enough to justify our 1D approach. 

 We assume that the secondary star is always disrupted by the H-free core of the primary during the merger in all our models. Although we do not consider the angular momentum injected by the secondary as it spirals-in through the CE, we do consider the envelope spin-up caused by the accretion of the secondary. 
 
For a first-order calculation for the maximum depth inside the He core that the secondary stream can penetrate, M$_\textrm{r,max}$, we assume that the secondary stream ballistically hits the He core with a free-fall velocity ($v_\textrm{st}$) and ram pressure (P$_\textrm{st}$) calculated as: 
 \begin{align}
 	P_\textrm{st} = \frac{1}{2} \rho_\textrm{st}v_\textrm{st}^{2}, \textrm{where}\,\, v_\textrm{st}= \sqrt{\cfrac{GM_\textrm{H-free core}} {R_\textrm{H-free core}}}
\label{eq:1}
 \end{align}
In the absence of shocks, we assume that the secondary star material conserves its entropy when it is disrupted into a stream and approaches the core.  In this case, the equation of state can be written as $P\propto\rho^\gamma$, where $\gamma$ is the adiabatic index. By assuming the stream is in lateral pressure equilibrium with the surrounding gas, we can estimate the stream density as:
 \begin{align}
  \rho_\textrm{st} &= \rho_\textrm{2,central}\Biggl(\frac{P_\textrm{H-free core}}{P_\textrm{2,central}}\Biggl)^{\frac{1}{\gamma_\textrm{H-free core}}}
 \label{eq:2}
 \end{align}

where, $P_\textrm{2,central}$ and $\rho_\textrm{2,central}$ are the central pressure and density of the secondary star. We use these quantities since the entropy at the centre of the secondary is the lowest and hence has the highest chance to penetrate the He core.  While the adiabatic index of the stream may change as it changes pressure, we use $\gamma_\textrm{H-free core}$ (its initial value as it enters the He core)  for simplicity and dimensional consistency. The location where $P_\textrm{st}$ equals the pressure inside the He core corresponds close to the lowest mass-coordinate  that the stream can penetrate (M$_\textrm{r,max}$). In reality, the stream could be shocked or could ignite explosive nuclear burning while penetrating the core, dramatically increasing its entropy before penetrating deep into the core. For this reason, we expect that our formula for the penetration depth corresponds to a conservative upper limit.

\section{Extended data}
\begin{landscape}
\fontsize{9}{12}\selectfont
\setlength\tabcolsep{8pt}
    \centering 
\label{tab:abundances}              
\begin{xltabular}{\textwidth}{llrcccccccccc}
\caption{Stellar parameters and abundances from FASTWIND analysis. Y$_{\rm He}$ is the helium to hydrogen ratio by number, 
V$_{\rm t}$ is the adopted microturbulent velocity, and spectral types are taken 
from \cite{fitzpatrick1991} and references in section 3.2. Also listed are the measured projected rotational velocities,
vsini, for the FEROS sample and, for convenience, the values published for the VFTS and FSMS samples \cite{mcevoy2015,hunter2009}.  } \\
\hline
Name & Spectral type	   & vsini & logL/L$_{\odot}$ &  T$_{\rm eff}$  & log$g$ & Y$_{\rm He}$  & V$_{\rm t}$ &  [C/H]  &  [N/H] & [O/H] \\   
 &   & (km\,s$^{-1}$) &  &  (K) &  & & (km\,s$^{-1}$) &   &   &  \\   
\hline       
\multicolumn{11}{c}{FEROS} \\
\hline  
    Sk-66 1 & B2Ia	& 51 & 5.67$\pm$0.03  & $19862^{+479}_{-502}$ & $2.40^{+0.08}_{-0.09}$  & $0.08^{+0.02}_{-0.00}$  & $15.5^{+2.4}_{-1.8}$  & $7.46^{+0.12}_{-0.15}$  & $8.10^{+0.09}_{-0.11}$  & $8.37^{+0.10}_{-0.08}$  \\  
    Sk-66 5 & B2.5Ia	&  9 & 5.77$\pm$0.04  & $17224^{+294}_{-297}$ & $2.09^{+0.06}_{-0.07}$  & $0.08^{+0.01}_{-0.00}$  & $15.1^{+2.3}_{-1.8}$  & $7.62^{+0.13}_{-0.11}$  & $8.16^{+0.09}_{-0.15}$  & $8.65^{+0.08}_{-0.11}$  \\  
   Sk-66 35 & BC1Ia	& 23 & 5.74$\pm$0.03  & $22214^{+686}_{-939}$ & $2.67^{+0.11}_{-0.13}$  & $0.13^{+0.03}_{-0.03}$  & $10.3^{+2.8}_{-2.2}$  & $7.98^{+0.18}_{-0.15}$  & $7.42^{+0.18}_{-0.20}$  & $8.61^{+0.13}_{-0.09}$  \\  
  Sk-66 106 & B1.5Ia	& 49 & 5.66$\pm$0.03  & $21730^{+718}_{-758}$ & $2.65^{+0.12}_{-0.11}$  & $0.08^{+0.03}_{-0.00}$  & $20.4^{+4.0}_{-1.9}$  & $7.26^{+0.20}_{-0.14}$  & $8.03^{+0.09}_{-0.10}$  & $8.29^{+0.09}_{-0.08}$  \\  
  Sk-66 118 & B2Iab	&  8 & 5.54$\pm$0.03  & $20133^{+591}_{-677}$ & $2.47^{+0.11}_{-0.11}$  & $0.18^{+0.04}_{-0.03}$  & $16.5^{+2.5}_{-2.6}$  & $7.61^{+0.18}_{-0.15}$  & $8.26^{+0.13}_{-0.09}$  & $8.33^{+0.15}_{-0.09}$  \\  
   Sk-67 14 & B1.5Ia	& 44 & 5.72$\pm$0.03  &$21245^{+1003}_{-808}$ & $2.59^{+0.14}_{-0.12}$  & $0.13^{+0.03}_{-0.05}$  & $21.2^{+3.1}_{-3.6}$  & $7.66^{+0.12}_{-0.20}$  & $8.22^{+0.11}_{-0.08}$  & $8.19^{+0.11}_{-0.10}$  \\  
   Sk-67 28 & B0.7Ia	& 43 & 5.50$\pm$0.03  & $24645^{+509}_{-553}$ & $2.73^{+0.08}_{-0.11}$  & $0.14^{+0.03}_{-0.04}$  & $19.1^{+5.5}_{-3.8}$  & $7.38^{+0.13}_{-0.28}$  & $8.00^{+0.13}_{-0.16}$  & $8.41^{+0.10}_{-0.11}$  \\  
   Sk-67 78 & B3Ia	&  9 & 5.49$\pm$0.03  & $16638^{+435}_{-260}$ & $2.10^{+0.07}_{-0.07}$  & $0.08^{+0.00}_{-0.00}$  & $15.0^{+1.9}_{-2.0}$  & $7.73^{+0.09}_{-0.12}$  & $8.01^{+0.17}_{-0.11}$  & $8.34^{+0.14}_{-0.07}$  \\  
   Sk-67 90 & B1.5Ia	& 53 & 5.81$\pm$0.03  & $21446^{+476}_{-651}$ & $2.51^{+0.08}_{-0.09}$  & $0.12^{+0.03}_{-0.03}$  & $20.4^{+1.9}_{-2.6}$  & $7.67^{+0.12}_{-0.13}$  & $8.14^{+0.10}_{-0.07}$  & $8.22^{+0.08}_{-0.08}$  \\  
  Sk-67 112 & B0.5Ia	& 49 & 5.62$\pm$0.03  & $25573^{+718}_{-415}$ & $2.82^{+0.11}_{-0.09}$  & $0.13^{+0.02}_{-0.04}$  & $27.5^{+2.0}_{-4.4}$  & $7.40^{+0.20}_{-0.20}$  & $7.84^{+0.16}_{-0.18}$  & $8.22^{+0.11}_{-0.12}$  \\  
  Sk-67 150 & B1.5Ia	& 28 & 5.50$\pm$0.04  & $23246^{+521}_{-716}$ & $2.66^{+0.10}_{-0.11}$  & $0.18^{+0.04}_{-0.03}$  & $23.3^{+3.0}_{-4.0}$  & $7.42^{+0.17}_{-0.22}$  & $8.23^{+0.10}_{-0.09}$  & $8.24^{+0.10}_{-0.07}$  \\  
  Sk-67 169 & B1Ia	& 26 & 5.51$\pm$0.03  & $23798^{+566}_{-552}$ & $2.73^{+0.10}_{-0.09}$  & $0.10^{+0.03}_{-0.02}$  & $19.1^{+2.8}_{-2.7}$  & $7.83^{+0.15}_{-0.10}$  & $7.78^{+0.10}_{-0.19}$  & $8.35^{+0.10}_{-0.09}$  \\  
  SK-67 172 & B2.5Iab	& 15 & 5.47$\pm$0.03  & $19331^{+443}_{-449}$ & $2.53^{+0.08}_{-0.09}$  & $0.08^{+0.02}_{-0.00}$  & $16.2^{+1.6}_{-1.9}$  & $7.67^{+0.11}_{-0.08}$  & $7.95^{+0.10}_{-0.10}$  & $8.33^{+0.09}_{-0.09}$  \\  
  Sk-67 173 & B0Ia	& 53 & 5.70$\pm$0.03  & $27480^{+510}_{-670}$ & $2.95^{+0.09}_{-0.10}$  & $0.09^{+0.02}_{-0.02}$  & $25.0^{+3.7}_{-3.0}$  & $7.35^{+0.18}_{-0.20}$  & $7.69^{+0.18}_{-0.20}$  & $8.53^{+0.13}_{-0.17}$  \\  
  Sk-67 206 & BN0.5Ia	& 59 & 5.61$\pm$0.02  & $25067^{+609}_{-438}$ & $2.73^{+0.11}_{-0.08}$  & $0.19^{+0.03}_{-0.04}$  & $15.1^{+4.0}_{-2.4}$  & $7.48^{+0.22}_{-0.28}$  & $8.07^{+0.16}_{-0.14}$  & $8.35^{+0.09}_{-0.15}$  \\  
  Sk-67 256 & BC1Ia	& 47 & 5.66$\pm$0.05  & $21993^{+545}_{-603}$ & $2.59^{+0.10}_{-0.10}$  & $0.10^{+0.03}_{-0.02}$  & $14.8^{+2.5}_{-2.5}$  & $7.98^{+0.12}_{-0.12}$  & $7.37^{+0.20}_{-0.16}$  & $8.58^{+0.11}_{-0.07}$  \\  
   Sk-68 26 & BC2Ia	& 48 & 5.71$\pm$0.02  & $17765^{+328}_{-381}$ & $2.14^{+0.07}_{-0.08}$  & $0.08^{+0.03}_{-0.00}$  & $13.0^{+2.9}_{-2.3}$  & $7.80^{+0.10}_{-0.14}$  & $7.23^{+0.26}_{-0.17}$  & $8.49^{+0.10}_{-0.08}$  \\  
   Sk-68 41 & B0.5Ia	& 67 & 5.66$\pm$0.06  & $25158^{+405}_{-540}$ & $2.76^{+0.08}_{-0.09}$  & $0.11^{+0.02}_{-0.03}$  & $22.4^{+3.9}_{-4.0}$  & $7.35^{+0.21}_{-0.18}$  & $7.84^{+0.12}_{-0.18}$  & $8.59^{+0.12}_{-0.09}$  \\  
   Sk-68 45 & B0Ia	& 60 & 5.65$\pm$0.03  & $27545^{+644}_{-475}$ & $2.90^{+0.09}_{-0.08}$  & $0.08^{+0.02}_{-0.00}$  & $25.1^{+3.5}_{-2.9}$  & $7.76^{+0.19}_{-0.13}$  & $7.43^{+0.19}_{-0.36}$  & $8.57^{+0.12}_{-0.15}$  \\  
   Sk-68 92 & B1.5Ia	& 42 & 5.62$\pm$0.02  & $21238^{+765}_{-765}$ & $2.62^{+0.11}_{-0.13}$  & $0.08^{+0.02}_{-0.00}$  & $22.8^{+2.9}_{-3.0}$  & $7.52^{+0.12}_{-0.21}$  & $8.00^{+0.08}_{-0.10}$  & $8.37^{+0.09}_{-0.09}$  \\  
  Sk-68 171 & B0.7Ia	& 53 & 5.62$\pm$0.03  & $24147^{+695}_{-442}$ & $2.73^{+0.11}_{-0.10}$  & $0.15^{+0.04}_{-0.03}$  & $28.4^{+1.6}_{-5.5}$  & $7.60^{+0.20}_{-0.20}$  & $8.02^{+0.09}_{-0.19}$  & $8.57^{+0.09}_{-0.10}$  \\  
   Sk-69 43 & B1Ia	& 42 & 5.58$\pm$0.04  & $23419^{+706}_{-382}$ & $2.74^{+0.12}_{-0.07}$  & $0.10^{+0.01}_{-0.03}$  & $24.9^{+2.2}_{-3.3}$  & $7.80^{+0.08}_{-0.14}$  & $8.02^{+0.08}_{-0.10}$  & $8.36^{+0.09}_{-0.05}$  \\  
   Sk-69 89 & B2.5Ia	&  9 & 5.62$\pm$0.03  & $17959^{+445}_{-337}$ & $2.23^{+0.08}_{-0.08}$  & $0.09^{+0.04}_{-0.02}$  & $20.0^{+2.2}_{-2.5}$  & $7.72^{+0.12}_{-0.12}$  & $8.41^{+0.11}_{-0.09}$  & $8.31^{+0.09}_{-0.10}$  \\  
  Sk-69 214 & B0.7Ia	& 41 & 5.68$\pm$0.02  & $24507^{+533}_{-697}$ & $2.96^{+0.10}_{-0.15}$  & $0.12^{+0.02}_{-0.03}$  & $23.3^{+3.2}_{-3.0}$  & $7.82^{+0.15}_{-0.19}$  & $7.92^{+0.13}_{-0.11}$  & $8.24^{+0.07}_{-0.10}$  \\  
  Sk-69 237 & B1IaNstr  & 49 & 5.68$\pm$0.02  & $24568^{+486}_{-657}$ & $2.90^{+0.10}_{-0.10}$  & $0.10^{+0.01}_{-0.03}$  & $21.6^{+2.4}_{-2.5}$  & $7.48^{+0.21}_{-0.17}$  & $8.08^{+0.09}_{-0.10}$  & $8.14^{+0.06}_{-0.10}$  \\  
  Sk-69 270 & B2.5Ia	& 10 & 5.72$\pm$0.03  & $16858^{+257}_{-334}$ & $2.06^{+0.06}_{-0.08}$  & $0.08^{+0.02}_{-0.00}$  & $14.0^{+1.7}_{-2.6}$  & $7.64^{+0.11}_{-0.16}$  & $8.32^{+0.14}_{-0.11}$  & $8.60^{+0.09}_{-0.11}$  \\  
  Sk-69 274 & B2.5Ia	&  8 & 5.74$\pm$0.03  & $18104^{+326}_{-404}$ & $2.22^{+0.08}_{-0.07}$  & $0.08^{+0.04}_{-0.01}$  & $16.3^{+1.4}_{-2.9}$  & $7.83^{+0.13}_{-0.10}$  & $8.41^{+0.09}_{-0.14}$  & $8.35^{+0.11}_{-0.08}$  \\  
  Sk-69 228 & BC1.5Ia	& 66 & 5.55$\pm$0.03  & $21020^{+811}_{-674}$ & $2.51^{+0.12}_{-0.11}$  & $0.08^{+0.03}_{-0.00}$  & $23.0^{+3.4}_{-3.1}$  & $7.91^{+0.16}_{-0.16}$  & $7.15^{+0.18}_{-0.16}$  & $8.25^{+0.10}_{-0.13}$  \\  
   Sk-70 78 & B0.7Ia	& 40 & 5.83$\pm$0.02  & $23483^{+632}_{-645}$ & $2.73^{+0.11}_{-0.12}$  & $0.08^{+0.03}_{-0.00}$  & $24.1^{+3.4}_{-3.1}$  & $7.58^{+0.18}_{-0.20}$  & $7.81^{+0.11}_{-0.16}$  & $8.39^{+0.09}_{-0.07}$  \\  
  SK-70 111 & B0.5Ia	& 10 & 5.80$\pm$0.05  & $26014^{+481}_{-476}$ & $2.93^{+0.09}_{-0.08}$  & $0.14^{+0.03}_{-0.02}$  & $25.7^{+1.9}_{-2.7}$  & $7.30^{+0.20}_{-0.13}$  & $7.95^{+0.12}_{-0.09}$  & $8.15^{+0.10}_{-0.07}$  \\  
  Sk-70 120 & B1Ia	& 36 & 5.69$\pm$0.02  & $22693^{+472}_{-730}$ & $2.58^{+0.09}_{-0.10}$  & $0.08^{+0.03}_{-0.00}$  & $21.7^{+3.0}_{-2.9}$  & $7.64^{+0.11}_{-0.18}$  & $7.81^{+0.09}_{-0.15}$  & $8.36^{+0.08}_{-0.08}$  \\  
   Sk-71 42 & B2Ia	& 43 & 5.83$\pm$0.02  & $19522^{+486}_{-534}$ & $2.41^{+0.09}_{-0.10}$  & $0.09^{+0.04}_{-0.01}$  & $18.9^{+2.2}_{-2.4}$  & $7.54^{+0.10}_{-0.19}$  & $8.34^{+0.12}_{-0.08}$  & $8.32^{+0.09}_{-0.10}$  \\
\hline       
\multicolumn{11}{c}{VFTS} \\
\hline  
    VFTS003 & B1Ia+	& 48 & 5.91$\pm$0.05  & $22546^{+513}_{-436}$ & $2.64^{+0.09}_{-0.07}$  & $0.08^{+0.03}_{-0.00}$  & $20.5^{+2.1}_{-2.8}$  & $7.15^{+0.31}_{-0.07}$  & $8.01^{+0.11}_{-0.13}$  & $8.29^{+0.09}_{-0.08}$  \\  
    VFTS028 & B0.7IaNwk & 50 & 5.73$\pm$0.10  & $24495^{+584}_{-332}$ & $2.67^{+0.09}_{-0.07}$  & $0.08^{+0.03}_{-0.00}$  & $17.4^{+2.7}_{-2.2}$  & $7.42^{+0.19}_{-0.23}$  & $7.54^{+0.16}_{-0.27}$  & $8.47^{+0.12}_{-0.09}$  \\  
    VFTS069 & B0.7Ib-Ia & 46 & 5.61$\pm$0.09  & $25072^{+325}_{-559}$ & $2.83^{+0.08}_{-0.10}$  & $0.13^{+0.05}_{-0.03}$  & $18.8^{+3.0}_{-4.4}$  & $7.28^{+0.22}_{-0.16}$  & $7.96^{+0.21}_{-0.15}$  & $8.35^{+0.10}_{-0.10}$  \\  
    VFTS082 & B0.5Ib-Iab& 49 & 5.28$\pm$0.12  & $26713^{+596}_{-402}$ & $3.02^{+0.11}_{-0.07}$  & $0.11^{+0.05}_{-0.02}$  & $15.9^{+3.3}_{-4.1}$  & $7.44^{+0.17}_{-0.27}$  & $8.08^{+0.20}_{-0.19}$  & $8.32^{+0.08}_{-0.11}$  \\  
    VFTS232 & B3Ia	& 42 & 5.01$\pm$0.06  & $17277^{+326}_{-278}$ & $2.48^{+0.07}_{-0.05}$  & $0.08^{+0.01}_{-0.00}$  & $ 5.0^{+1.9}_{-0.0}$  & $7.28^{+0.10}_{-0.09}$  & $7.66^{+0.09}_{-0.10}$  & $8.20^{+0.10}_{-0.10}$  \\  
    VFTS302 & B1.5Ib	& 34 & 4.90$\pm$0.07  & $23063^{+517}_{-709}$ & $3.14^{+0.02}_{-0.16}$  & $0.08^{+0.04}_{-0.00}$  & $10.2^{+1.6}_{-2.9}$  & $7.54^{+0.15}_{-0.15}$  & $7.87^{+0.21}_{-0.13}$  & $8.42^{+0.09}_{-0.06}$  \\  
    VFTS315 & B1Ib	& 31 & 4.73$\pm$0.08  & $24899^{+485}_{-772}$ & $3.28^{+0.03}_{-0.09}$  & $0.10^{+0.02}_{-0.02}$  & $11.4^{+1.6}_{-2.2}$  & $7.41^{+0.20}_{-0.16}$  & $7.49^{+0.14}_{-0.25}$  & $8.58^{+0.05}_{-0.09}$  \\  
    VFTS431 & B1.5IaNstr & 41 & 5.80$\pm$0.10  & $20824^{+499}_{-611}$ & $2.43^{+0.10}_{-0.10}$  & $0.15^{+0.03}_{-0.04}$  & $14.8^{+1.4}_{-2.8}$  & $7.53^{+0.15}_{-0.18}$  & $8.31^{+0.14}_{-0.11}$  & $8.53^{+0.08}_{-0.12}$  \\  
    VFTS533 & B1.5Ia+pNwk& 57 & 5.92$\pm$0.07  & $19210^{+579}_{-429}$ & $2.31^{+0.12}_{-0.08}$  & $0.09^{+0.03}_{-0.01}$  & $15.0^{+2.6}_{-1.7}$  & $7.40^{+0.17}_{-0.21}$  & $7.65^{+0.15}_{-0.20}$  & $8.66^{+0.10}_{-0.10}$  \\	 
    VFTS590 & B0.7Iab   & 60 & 5.85$\pm$0.07  & $25318^{+397}_{-370}$ & $2.72^{+0.07}_{-0.07}$  & $0.12^{+0.05}_{-0.03}$  & $14.6^{+5.1}_{-2.1}$  & $7.23^{+0.23}_{-0.13}$  & $8.16^{+0.17}_{-0.17}$  & $8.15^{+0.11}_{-0.12}$  \\  
    VFTS696 & B0.7Ib-abNwk& 53 & 5.61$\pm$0.08  & $24628^{+578}_{-414}$ & $2.76^{+0.09}_{-0.08}$  & $0.08^{+0.03}_{-0.00}$  & $15.1^{+3.2}_{-2.1}$  & $7.36^{+0.21}_{-0.20}$  & $7.49^{+0.15}_{-0.24}$  & $8.51^{+0.09}_{-0.10}$  \\  
    VFTS732 & B1.5IapNwk& 45 & 5.61$\pm$0.08  & $22404^{+492}_{-059}$ & $2.69^{+0.10}_{-0.15}$  & $0.08^{+0.04}_{-0.00}$  & $13.5^{+2.1}_{-2.4}$  & $7.45^{+0.20}_{-0.22}$  & $7.39^{+0.15}_{-0.24}$  & $8.46^{+0.09}_{-0.09}$  \\  
    VFTS831 & B5Ia	& 41 & 5.25$\pm$0.06  & $16220^{+215}_{-122}$ & $2.17^{+0.05}_{-0.05}$  & $0.08^{+0.03}_{-0.00}$  & $ 8.5^{+1.1}_{-1.3}$  & $7.45^{+0.09}_{-0.08}$  & $7.78^{+0.23}_{-0.35}$  & $8.45^{+0.12}_{-0.10}$  \\  
    VFTS867 & B1IbNwk	& 32 & 4.99$\pm$0.07  & $26274^{+248}_{-471}$ & $3.37^{+0.01}_{-0.09}$  & $0.09^{+0.03}_{-0.01}$  & $13.2^{+2.1}_{-1.4}$  & $7.49^{+0.13}_{-0.25}$  & $7.27^{+0.17}_{-0.17}$  & $8.54^{+0.07}_{-0.08}$  \\  
\hline       
\multicolumn{11}{c}{FSMS} \\
\hline  									   
    N11-001 & B2Ia	& 52 & 5.81$\pm$0.03  & $18594^{+527}_{-882}$ & $2.27^{+0.09}_{-0.13}$  & $0.08^{+0.04}_{-0.00}$  & $15.5^{+2.7}_{-2.3}$  & $7.53^{+0.17}_{-0.15}$  & $8.26^{+0.13}_{-0.14}$  & $8.29^{+0.16}_{-0.13}$  \\  
    N11-002 & B3Ia	& 53 & 5.45$\pm$0.03  & $17035^{+568}_{-343}$ & $2.21^{+0.09}_{-0.09}$  & $0.08^{+0.04}_{-0.00}$  & $12.4^{+2.3}_{-2.2}$  & $7.82^{+0.15}_{-0.14}$  & $8.14^{+0.19}_{-0.17}$  & $8.26^{+0.18}_{-0.16}$  \\  
    N11-008 & B0.5Ia	& 43 & 5.39$\pm$0.03  & $25540^{+682}_{-545}$ & $2.87^{+0.11}_{-0.11}$  & $0.09^{+0.05}_{-0.02}$  & $16.8^{+5.9}_{-3.5}$  & $7.52^{+0.19}_{-0.16}$  & $7.75^{+0.16}_{-0.18}$  & $8.31^{+0.10}_{-0.11}$  \\  
    N11-015 & B0.7Ib	& 58 & 5.30$\pm$0.03  & $26082^{+665}_{-619}$ & $3.03^{+0.13}_{-0.09}$  & $0.15^{+0.06}_{-0.03}$  & $ 9.6^{+2.8}_{-2.4}$  & $7.72^{+0.19}_{-0.17}$  & $7.52^{+0.21}_{-0.35}$  & $8.56^{+0.14}_{-0.10}$  \\ 
    N11-016 & B1Ib	& 54 & 5.19$\pm$0.03  & $22832^{+876}_{-671}$ & $2.74^{+0.12}_{-0.10}$  & $0.08^{+0.05}_{-0.01}$  & $13.7^{+3.4}_{-2.0}$  & $7.67^{+0.17}_{-0.12}$  & $7.91^{+0.16}_{-0.14}$  & $8.37^{+0.11}_{-0.12}$  \\  
    N11-024 & B1Ib	& 30 & 5.05$\pm$0.03  & $23671^{+719}_{-986}$ & $2.91^{+0.13}_{-0.14}$  & $0.13^{+0.03}_{-0.04}$  & $11.0^{+3.2}_{-2.8}$  & $7.65^{+0.16}_{-0.15}$  & $8.02^{+0.16}_{-0.16}$  & $8.32^{+0.12}_{-0.11}$  \\  
    N11-036 & B0.5Ib	& 30 & 5.06$\pm$0.16  & $25118^{+777}_{-690}$ & $3.10^{+0.14}_{-0.11}$  & $0.08^{+0.05}_{-0.01}$  & $11.1^{+2.6}_{-3.1}$  & $7.59^{+0.12}_{-0.17}$  & $7.85^{+0.16}_{-0.15}$  & $8.32^{+0.14}_{-0.09}$  \\  
    N11-054 & B1Ib	& 45 & 5.14$\pm$0.14  & $25600^{+877}_{-811}$ & $3.15^{+0.12}_{-0.14}$  & $0.08^{+0.04}_{-0.00}$  & $ 8.2^{+2.4}_{-2.6}$  & $7.80^{+0.12}_{-0.17}$  & $7.10^{+0.20}_{-0.20}$  & $8.55^{+0.08}_{-0.13}$  \\  
NGC2004-010 & B2.5Iab	& 30 & 5.16$\pm$0.03  & $18038^{+490}_{-537}$ & $2.45^{+0.09}_{-0.11}$  & $0.08^{+0.05}_{-0.01}$  & $12.5^{+2.6}_{-2.6}$  & $7.63^{+0.11}_{-0.14}$  & $8.18^{+0.13}_{-0.22}$  & $8.19^{+0.19}_{-0.15}$  \\  
NGC2004-011 & B1.5Ia	& 40 & 5.23$\pm$0.03  & $22531^{+982}_{-785}$ & $2.73^{+0.14}_{-0.12}$  & $0.12^{+0.05}_{-0.03}$  & $13.8^{+3.1}_{-3.2}$  & $7.67^{+0.14}_{-0.15}$  & $7.83^{+0.19}_{-0.15}$  & $8.38^{+0.15}_{-0.12}$  \\  
NGC2004-012 & B1.5Iab	& 40 & 4.99$\pm$0.05  & $23801^{+845}_{-855}$ & $2.99^{+0.14}_{-0.13}$  & $0.12^{+0.03}_{-0.05}$  & $ 9.8^{+2.6}_{-3.1}$  & $7.75^{+0.15}_{-0.17}$  & $7.89^{+0.19}_{-0.14}$  & $8.42^{+0.11}_{-0.12}$  \\  
NGC2004-021 & B1.5Ib	& 40 & 4.83$\pm$0.06  & $22795^{+960}_{-724}$ & $2.97^{+0.12}_{-0.13}$  & $0.10^{+0.04}_{-0.03}$  & $ 9.3^{+2.8}_{-2.9}$  & $7.91^{+0.16}_{-0.15}$  & $7.34^{+0.20}_{-0.20}$  & $8.56^{+0.11}_{-0.12}$  \\  
NGC2004-022 & B1.5Ib	& 30 & 4.77$\pm$0.03  & $23928^{+928}_{-890}$ & $3.19^{+0.14}_{-0.14}$  & $0.08^{+0.04}_{-0.00}$  & $ 8.4^{+2.7}_{-1.8}$  & $7.65^{+0.13}_{-0.18}$  & $7.81^{+0.15}_{-0.17}$  & $8.48^{+0.11}_{-0.12}$  \\ 
\hline
\end{xltabular}
\end{landscape}

\begin{landscape}
\fontsize{9}{11}\selectfont
\setlength\tabcolsep{1.5pt}
    \centering 
\label{tab:abundances}              
\begin{xltabular}{\textwidth}{lcccccccccccccccccc}
\caption{Parameters of the Case\,B merger models: M$_\textrm{1,i}$,  M$_\textrm{2,i}$ and P$_\textrm{i}$ are the initial parameters of the binary system. M$_\textrm{1,merger}$ is the mass of the primary at the time of the merger.  BSG parameters are averaged during the BSG phase of the models, and are as reported in Figs.~\ref{log_n_c_n_o} and \ref{surf_bsgs_all}. M$_\textrm{C-dep}$ is the mass at core C-depletion,  close to its expected value at explosion and SN prog. indicates the expected structure of the star at explosion:  `BSG' for progenitors with log\,T$\textrm{eff}\geq4.0$ and `RSG' for  progenitors which have log\,T$\textrm{eff}<4.0$. Where convergence was possible until core C-depletion, we have evolved a given binary system at both limits of core penetration.} \label{tab:caseB} \\
\hline
M$_\textrm{1,i}$ & M$_\textrm{2,i}$ & P$_\textrm{i}$ & f$_\textrm{c}$  &  M$_\textrm{1,merger}$  & M$_\textrm{BSG}$ & M$_\textrm{H-free core, BSG}$ & t$_\textrm{BSG}$ & log\,g$_\textrm{BSG}$ & log\,$\left(\displaystyle\frac{L_\textrm{BSG}}{L_{\odot}}\right)$ & log\,T$_\textrm{eff,BSG}$ & $\epsilon_\textrm{C,BSG}$ & $\epsilon_\textrm{N,BSG}$ & $\epsilon_\textrm{O,BSG}$ & $\epsilon_\textrm{He,BSG}$  & M$_\textrm{C-dep}$ & SN prog. \\
\hline
(\Msun) & (\Msun) & (d) &  (\%) &  (\Msun) & (\Msun) & (\Msun)  &  (Myr) & (cm/s$^{2}$) & & (K)  &  &   &   & & (\Msun) &  \\
  \noalign{\global\arrayrulewidth=0.2mm}
\hline
10.0 & 1.0 & 24 & 0.0 & 10 & 10 & 2.2 & 2.0 & 2.2 & 4.4 & 4.0 & 7.6 & 7.9 & 8.3 & 0.14 & 10 & RSG \\
11.2 & 9.0 & 1413 & 0.06 & 8 & 17 & 2.7 & 1.4 & 3.4 & 5.0 & 4.4 & 7.4 & 8.2 & 8.2 & 0.17 & 17 & BSG \\
11.2 & 9.0 & 1413 & 0.0 & 8 & 17 & 2.8 & 1.3 & 3.5 & 5.0 & 4.5 & 7.4 & 8.2 & 8.2 & 0.17 & 17 & BSG \\
14.1 & 5.7 & 1884 & 0.1 & 11 & 17 & 3.6 & 1.0 & 3.3 & 5.0 & 4.4 & 7.4 & 8.2 & 8.2 & 0.15 & 17 & BSG \\
14.1 & 12.0 & 1995 & 0.07 & 10 & 22 & 3.6 & 1.0 & 3.2 & 5.3 & 4.4 & 7.4 & 8.3 & 8.2 & 0.2 & 22 & BSG \\
14.1 & 5.7 & 1884 & 0.0 & 11 & 17 & 4.0 & 0.9 & 3.3 & 4.9 & 4.4 & 7.4 & 8.1 & 8.2 & 0.14 & 17 & BSG \\
14.1 & 12.0 & 1995 & 0.0 & 10 & 22 & 4.0 & 0.9 & 3.1 & 5.3 & 4.4 & 7.4 & 8.3 & 8.2 & 0.2 & 22 & BSG \\
15.8 & 1.1 & 562 & 0.25 & 16 & 17 & 3.5 & 1.0 & 3.2 & 5.0 & 4.4 & 7.6 & 8.2 & 8.3 & 0.18 & 17 & BSG \\
15.8 & 1.9 & 596 & 0.2 & 16 & 18 & 3.7 & 1.0 & 3.1 & 5.1 & 4.4 & 7.6 & 8.1 & 8.3 & 0.17 & 18 & BSG \\
15.8 & 2.7 & 668 & 0.17 & 16 & 18 & 3.9 & 0.9 & 3.4 & 5.1 & 4.4 & 7.5 & 8.1 & 8.3 & 0.16 & 18 & BSG \\
15.8 & 10.3 & 2371 & 0.08 & 13 & 24 & 4.4 & 0.8 & 3.3 & 5.3 & 4.4 & 7.4 & 8.2 & 8.2 & 0.16 & 24 & BSG \\
15.8 & 1.6 & 47 & 0.0 & 16 & 16 & 3.4 & 1.0 & 2.6 & 5.0 & 4.2 & 7.6 & 8.1 & 8.3 & 0.18 & 16 & BSG \\
15.8 & 1.9 & 596 & 0.12 & 16 & 17 & 4.7 & 0.7 & 3.0 & 5.0 & 4.3 & 7.5 & 8.0 & 8.3 & 0.14 & 17 & BSG \\
15.8 & 2.7 & 668 & 0.0 & 16 & 18 & 3.3 & 1.1 & 3.6 & 5.1 & 4.5 & 7.6 & 8.1 & 8.3 & 0.18 & 18 & BSG \\
15.8 & 10.3 & 2371 & 0.0 & 13 & 24 & 4.8 & 0.7 & 3.2 & 5.3 & 4.4 & 7.4 & 8.2 & 8.2 & 0.15 & 24 & BSG \\
15.8 & 1.1 & 562 & 0.0 & 16 & 17 & 4.7 & 0.7 & 2.9 & 5.0 & 4.3 & 7.6 & 8.0 & 8.3 & 0.15 & 16 & BSG \\
15.8 & 1.1 & 1585 & 0.0 & 16 & 17 & 4.7 & 0.7 & 2.9 & 5.0 & 4.3 & 7.6 & 8.0 & 8.3 & 0.14 & 16 & BSG \\
15.8 & 1.6 & 47 & 0.0 & 16 & 16 & 4.6 & 0.8 & 2.3 & 5.0 & 4.2 & 7.6 & 8.0 & 8.3 & 0.14 & 16 & BSG \\
17.8 & 8.9 & 2661 & 0.09 & 16 & 25 & 5.2 & 0.7 & 3.2 & 5.3 & 4.4 & 7.4 & 8.2 & 8.2 & 0.15 & 25 & BSG \\
17.8 & 8.9 & 2661 & 0.0 & 16 & 25 & 5.7 & 0.6 & 3.1 & 5.3 & 4.4 & 7.4 & 8.1 & 8.2 & 0.14 & 25 & BSG \\
17.8 & 1.2 & 631 & 0.0 & 18 & 18 & 5.6 & 0.6 & 2.9 & 5.1 & 4.3 & 7.6 & 8.0 & 8.3 & 0.15 & 18 & BSG \\
17.8 & 1.2 & 1995 & 0.0 & 18 & 18 & 5.7 & 0.6 & 2.8 & 5.1 & 4.3 & 7.6 & 8.0 & 8.3 & 0.14 & 18 & BSG \\
20.0 & 11.0 & 2985 & 0.09 & 18 & 29 & 6.0 & 0.6 & 3.1 & 5.5 & 4.4 & 7.4 & 8.2 & 8.2 & 0.16 & 29 & BSG \\
20.0 & 11.0 & 2985 & 0.0 & 18 & 29 & 6.7 & 0.6 & 3.0 & 5.5 & 4.4 & 7.4 & 8.2 & 8.2 & 0.15 & 29 & BSG \\
20.0 & 15.0 & 3162 & 0.08 & 18 & 33 & 6.1 & 0.6 & 3.2 & 5.6 & 4.5 & 7.4 & 8.2 & 8.2 & 0.19 & 33 & BSG \\
20.0 & 15.0 & 3162 & 0.0 & 18 & 33 & 6.7 & 0.5 & 3.1 & 5.6 & 4.4 & 7.4 & 8.2 & 8.2 & 0.18 & 33 & BSG \\
20.0 & 1.0 & 708 & 0.0 & 20 & 20 & 6.6 & 0.6 & 2.7 & 5.2 & 4.3 & 7.6 & 8.1 & 8.3 & 0.15 & 20 & BSG \\
20.0 & 1.0 & 2113 & 0.0 & 20 & 20 & 6.7 & 0.6 & 2.6 & 5.2 & 4.3 & 7.6 & 8.1 & 8.3 & 0.15 & 20 & BSG \\
20.0 & 1.4 & 750 & 0.0 & 20 & 20 & 6.6 & 0.6 & 2.8 & 5.2 & 4.3 & 7.6 & 8.1 & 8.3 & 0.15 & 20 & BSG \\
20.0 & 1.4 & 2239 & 0.0 & 20 & 21 & 6.7 & 0.6 & 2.7 & 5.2 & 4.3 & 7.6 & 8.1 & 8.3 & 0.15 & 20 & BSG \\
20.0 & 2.0 & 16 & 0.0 & 20 & 20 & 6.5 & 0.6 & 2.5 & 5.2 & 4.2 & 7.6 & 7.9 & 8.3 & 0.14 & 20 & BSG \\
25.1 & 1.3 & 1000 & 0.0 & 24 & 25 & 9.2 & 0.5 & 2.2 & 5.4 & 4.2 & 7.6 & 8.1 & 8.3 & 0.16 & 25 & BSG \\
25.1 & 1.3 & 2661 & 0.0 & 24 & 25 & 9.3 & 0.5 & 2.4 & 5.5 & 4.3 & 7.5 & 8.1 & 8.3 & 0.16 & 24 & BSG \\
31.6 & 3.2 & 27 & 0.25 & 30 & 30 & 8.9 & 0.5 & 2.7 & 5.6 & 4.4 & 7.6 & 8.2 & 8.2 & 0.23 & 30 & BSG \\
31.6 & 13.3 & 28 & 0.28 & 23 & 29 & 10.8 & 0.4 & 2.5 & 5.6 & 4.3 & 7.2 & 8.4 & 8.1 & 0.21 & 29 & BSG \\
31.6 & 13.3 & 28 & 0.0 & 23 & 28 & 12.6 & 0.4 & 2.4 & 5.6 & 4.3 & 7.2 & 8.3 & 8.1 & 0.17 & 27 & RSG \\
31.6 & 7.9 & 126 & 0.0 & 30 & 32 & 12.6 & 0.4 & 2.2 & 5.6 & 4.2 & 7.4 & 8.1 & 8.2 & 0.15 & 31 & RSG \\
31.6 & 3.2 & 27 & 0.0 & 30 & 30 & 12.4 & 0.4 & 2.4 & 5.6 & 4.3 & 7.5 & 8.1 & 8.2 & 0.16 & 28 & RSG \\
35.5 & 17.7 & 668 & 0.12 & 19 & 36 & 13.1 & 0.4 & 2.4 & 5.8 & 4.3 & 7.0 & 8.4 & 8.1 & 0.25 & 35 & BSG \\
35.5 & 17.7 & 668 & 0.0 & 19 & 35 & 14.7 & 0.4 & 2.1 & 5.8 & 4.2 & 7.0 & 8.4 & 8.1 & 0.21 & 34 & BSG \\
39.8 & 20.7 & 631 & 0.14 & 22 & 40 & 15.1 & 0.4 & 2.1 & 5.9 & 4.3 & 7.0 & 8.5 & 8.1 & 0.27 & 40 & BSG \\
39.8 & 31.8 & 376 & 0.0 & 21 & 50 & 16.8 & 0.3 & 2.3 & 6.1 & 4.3 & 7.2 & 8.6 & 8.1 & 0.38 & 46 & BSG \\
39.8 & 4.0 & 30 & 0.0 & 37 & 36 & 16.8 & 0.3 & 2.4 & 5.8 & 4.3 & 7.4 & 8.2 & 8.2 & 0.18 & 33 & RSG \\
39.8 & 10.0 & 38 & 0.0 & 37 & 36 & 17.0 & 0.3 & 2.3 & 5.8 & 4.3 & 7.3 & 8.2 & 8.2 & 0.16 & 33 & RSG \\
39.8 & 20.7 & 631 & 0.0 & 22 & 41 & 17.0 & 0.3 & 2.3 & 5.9 & 4.3 & 7.0 & 8.4 & 8.1 & 0.23 & 39 & RSG \\

\hline
\end{xltabular}
\end{landscape}

\section*{Data availability}
Observational sample and model parameters can be made available on request. Please contact AM for the same.

\section*{Code availability}
The code used to calculate the merger models can be shared on reasonable request. Please contact AM for the same. 

\clearpage


\maketitle

\bibliographystyle{naturemag}
\bibliography{master.bib}

\end{document}